\documentclass[fleqn,usenatbib]{mnras}

\usepackage[T1]{fontenc}

\DeclareRobustCommand{\VAN}[3]{#2}
\let\VANthebibliography\thebibliography
\def\thebibliography{\DeclareRobustCommand{\VAN}[3]{##3}\VANthebibliography}

\usepackage{graphicx}	
\usepackage{amsmath}	
\usepackage{amssymb}	
\usepackage{multirow}   
\usepackage{newtxtext,newtxmath}
\usepackage{soul}

\graphicspath{{figs/}}

\newcommand{\hst}{\textit{HST}}
\newcommand{\jwst}{\textit{JWST}}
\newcommand{\beagle}{\textsc{beagle}}
\newcommand{\prospector}{\textsc{Prospector}}
\newcommand{\Lya}{\ion{Ly}{$\alpha$}}
\newcommand{\Msun}{M$_\odot$}
\newcommand{\zsys}{\ensuremath{z_{\mathrm{sys}}}}
\newcommand{\larsonlae}{EGS\textunderscore z910\textunderscore 44164}

\title[Galaxy overdensities around $z = 8.7$ LAEs]{Insight from \textit{JWST}/NIRCam into galaxy overdensities around bright \ion{Ly}{$\alpha$} emitters during reionization: implications for ionized bubbles at $z \sim 9$}

\author[Whitler et al.]{Lily Whitler,$^{1}$\thanks{email: \href{mailto:lwhitler@arizona.edu}{lwhitler@arizona.edu}}\thanks{NSF Graduate Research Fellow}
Daniel P. Stark,$^{1}$
Ryan Endsley,$^{2}$
Zuyi Chen,$^{1}$
Charlotte Mason,$^{3,4}$
Michael W. Topping,$^{1}$
\newauthor
and St\'{e}phane Charlot$^{5}$
\\
$^{1}$Steward Observatory, University of Arizona, 933 N Cherry Ave, Tucson, AZ 85721, USA \\
$^{2}$Department of Astronomy, University of Texas, Austin, TX 78712, USA \\
$^{3}$Cosmic Dawn Center (DAWN) \\
$^{4}$Niels Bohr Institute, University of Copenhagen, Jagtvej 128, 2200 Copenhagen N, Denmark \\
$^{5}$Sorbonne Universit\'{e}, UPMC-CNRS, UMR7095, Institut d'Astrophysique de Paris, F-75014, Paris, France \\
}

\date{Accepted XXX. Received YYY; in original form ZZZ}

\pubyear{2024}

\begin{document}
\label{firstpage}
\pagerange{\pageref{firstpage}--\pageref{lastpage}}
\maketitle

\begin{abstract}
Several studies have detected Lyman-alpha (\Lya) from bright ($M_\textsc{uv}\lesssim-21.5$) galaxies during the early stages of reionization despite the significantly neutral intergalactic medium. To explain these detections, it has been suggested that $z>7$ \Lya\ emitters (LAEs) inhabit physical Mpc (pMpc)-scale ionized regions powered by overdensities of faint galaxies, but systematic searches for these overdensities near LAEs have been challenging. Here, we use CEERS \textit{JWST}/NIRCam imaging to search for large-scale galaxy overdensities near two very UV-bright, $z=8.7$ LAEs in the EGS field. We colour select 27 $z=8.4-9.1$ candidates, including the one LAE in the footprint (EGSY8p7). From SED models, we infer moderately faint UV luminosities ($-21.2\lesssim{M_\textsc{uv}}\lesssim -19.1$) and stellar masses of $M_*\approx10^{7.5-8.8}$\,\Msun. All are efficient ionizing agents ($\xi_{\text{ion}}^{*}\approx10^{25.5-26.0}$\,Hz\,erg$^{-1}$) and are generally morphologically simple with only one compact ($r_e\lesssim140$ to $\sim650$\,pc) star-forming component. 13 candidates lie within 5\,arcmin of EGSY8p7, leading to a factor-of-four galaxy overdensity at $\lesssim 5$\,arcmin ($\sim 1.4$\,projected\,pMpc at $z\sim8.7$) separations from EGSY8p7. Separations of $10-15$\,arcmin ($\sim2.7-4.1$\,projected\,pMpc) are consistent with an average field. The spatial distribution of our sample may qualitatively suggest an $R\geq2$\,pMpc ionized bubble encompassing both LAEs in EGS, which is theoretically unexpected but may be possible for a galaxy population $4\times$ more numerous than the average to create with moderate escape fractions ($f_\text{esc}\gtrsim0.15$) over long times ($\gtrsim200$\,Myr). Upcoming spectroscopic follow-up will characterize the size of any ionized bubble that may exist and the properties of the galaxies powering such a bubble.
\end{abstract}

\begin{keywords}
galaxies: high-redshift -- dark ages, reionization, first stars
\end{keywords}



\section{Introduction} \label{sec:intro}

During cosmic reionization, nearly all hydrogen atoms in the intergalactic medium (IGM) were ionized by radiation emerging from the earliest galaxies. Reionization is thus inextricably linked to the formation and evolution of these early galaxies, and significant effort over the past two decades has been dedicated to understanding this connection \citep[e.g.][]{madau1999, fan2006, robertson2015}.

A widely utilised observational probe of reionization is the Lyman-alpha (\Lya) emission from early galaxies. As a resonant line, \Lya\ emitted from galaxies during reionization is highly susceptible to scattering by neutral hydrogen in the IGM, and is therefore expected to become increasingly attenuated as the Universe becomes increasingly neutral \citep[e.g.][]{miralda-escude1998, haiman2002, malhotra2004, santos2004, mcquinn2007, mesinger2008, dijkstra2014, mason2018a}. Observationally, it has been shown that the fraction of typical rest-UV continuum selected galaxies observed to have strong \Lya\ (rest-frame equivalent width $\text{EW}_{\Lya} > 25$\,\AA) decreases rapidly at $z \gtrsim 6$ \citep[e.g.][]{stark2010, fontana2010, ono2012, treu2013, schenker2014, caruana2014, pentericci2018, jung2018, jung2020} and the luminosity function of narrowband-selected \Lya\ emitters (LAEs) declines between $z \sim 6$ and $z \sim 7$ \citep[e.g.][]{ouchi2010, zheng2017, ota2017, konno2018, itoh2018, hu2019, ouchi2020, goto2021, wold2022}. Altogether, this implies that the IGM transitions from being significantly ($\gtrsim 50$\,per\,cent) neutral at $z \gtrsim 7$ \citep[e.g.][]{mason2019_kmos, whitler2020, bolan2022} to primarily ionized by $z \sim 5 - 6$, a picture that is also consistent with other observational probes of the reionization timeline such as gamma ray bursts \citep[e.g.][]{totani2006, gallerani2008, hartoog2015}, the cosmic microwave background \citep[e.g.][]{planck2020}, and quasars \citep[e.g.][]{greig2017, greig2022, davies2018, wang2020, yang2020, zhu2022, jin2023}.

Recently, attention has turned to understanding the cause of \Lya\ emission observed from galaxies at $z > 7$ \citep[e.g.][]{vanzella2011, oesch2015, zitrin2015, castellano2016, castellano2018, endsley2021_LyA, tilvi2020, jung2020, jung2022, jung2023, larson2022, tang2023, bunker2023, saxena2023}. It is expected that \Lya\ emission from galaxies at these epochs is highly attenuated by the partially neutral IGM, so the detection of \Lya\ is somewhat unexpected. One possible explanation for strong observed \Lya\ is that these LAEs are embedded in large ionized bubbles; if a galaxy is surrounded by a large volume of ionized gas, the \Lya\ it emits can cosmologically redshift far into the damping wing before encountering intergalactic neutral hydrogen, thereby preserving much of the intrinsic emission escaping the galaxy \citep{wyithe2005, furlanetto2006, weinberger2018}. Such large ionized bubbles could result from significant overdensities of galaxies, where copious amounts of ionizing photons are produced by an unusually abundant galaxy population \citep[e.g.][]{barkana2004, furlanetto2004, iliev2006, dayal2018, qin2022}. However, it is also possible that \Lya\ may be detectable from galaxies in relatively small ionized regions if, for example, the majority of \Lya\ photons that emerge from the galaxy are already significantly redshifted from systemic \citep[as has frequently been observed for luminous sources; e.g.][]{erb2014, willott2015, mason2018b, hashimoto2019, matthee2020, endsley2022_velocityOffset} or the galaxy has a very intense radiation field that facilitates the production and visibility of \Lya\ \citep{stark2017, endsley2021_LyA, simmonds2023, tang2023, roberts-borsani2023}. Thus, it is unclear if galaxies observed to be emitting strong \Lya\ reside in small bubbles primarily powered by their own ionizing emission, or large bubbles generated by the combined ionizing flux of all neighbouring galaxies contributing to the reionization process.

Observations with the \textit{Hubble Space Telescope} (\hst) and ground-based facilities delivered the first insights into the nature of overdensities around UV-luminous LAEs at $z>7$. Using dedicated \hst/Wide Field Camera 3 (WFC3) imaging, \cite{leonova2022} found an excess of moderately UV-bright ($M_\textsc{uv} \lesssim -20$), photometrically selected galaxies surrounding three extremely UV-luminous ($M_\textsc{uv} \approx -22$) LAEs in the Extended Groth Strip (EGS) field at $z = 7.5$ \citep{roberts-borsani2016, stark2017}, $z = 7.7$ \citep{oesch2015}, and $z = 8.7$ \citep{zitrin2015} on projected scales of $R \sim 0.3$\,physical\,Mpc (pMpc) (i.e. the physical scale of the WFC3 field of view at these redshifts, $\sim 1$\,arcmin). Additional spectroscopic efforts with Keck/MOSFIRE have revealed \Lya\ emission from several moderately bright ($-20.5 < M_\textsc{uv} < -20.0$) galaxies $< 0.7$\,pMpc away from the extremely UV-luminous LAE at $z = 7.7$ (\citealt{tilvi2020}; see also \citealt{jung2022} and \textit{JWST}/Near Infrared Spectrograph observations by \citealt{jung2023}), and at $z = 8.7$, a second bright LAE only 4\,pMpc away from the already known LAE in the field \citep{zitrin2015} was confirmed with MOSFIRE \citep{larson2022}. Both of these discoveries lent additional evidence that these rare systems occupied significant galaxy overdensities associated with large ionized bubble. Furthermore, the two $z = 8.7$ LAEs are also accompanied by an overdensity of bright ($M_\textsc{uv} \lesssim -20.8$) \textit{HST}-selected photometric candidates over the entire EGS field \citep{finkelstein2022, larson2022} -- some of which are now spectroscopically confirmed to lie at $z = 8.7$ \citep{tang2023, nakajima2023, arrabalharo2023a} -- which may further suggest that these systems exist in a large ionized bubble.

\textit{HST} could identify signatures of overdensities of moderately bright galaxies relatively close to bright LAEs and larger scale overdensities of brighter galaxies. However, a key question moving forward is whether overdensities of faint galaxies regularly extend to large, $R \gtrsim 1$\,pMpc spatial scales. Significant large-scale overdensities associated with LAEs may be suggestive of large ionized bubbles being powered by many faint neighbours of LAEs, whereas a lack of such structures may be indicative of smaller ionized bubbles and physics internal to the LAEs facilitating \Lya\ escape (e.g. outflows leading to large \Lya\ velocity offsets). To help distinguish these scenarios, it will thus be important to (1) map faint galaxies in regions surrounding extremely UV-luminous LAEs to separations of a few pMpc and (2) obtain constraints on the density and ionizing properties of these faint ($M_\textsc{uv} \gtrsim -20$) neighbouring galaxies. Fortunately, \textit{JWST} enables both of these observational studies. First, \textit{JWST} allows for the photometric identification of faint neighbouring galaxies of LAEs, placing the first constraints on their spatial distribution and ionizing properties. Then, once faint neighbours have been identified, spectroscopic follow up targeting \Lya\ can begin to characterize the sizes and morphologies of ionized regions in the vicinity of the bright LAEs.

In this work, we characterize the overdensity potentially associated with two of the highest redshift known, extremely UV-bright LAEs, EGSY8p7 \citep{zitrin2015} and \larsonlae\ \citep{larson2022}, which lie in the EGS field at $z = 8.7$. We use deep \jwst/Near Infrared Camera \citep[NIRCam;][]{rieke2023} imaging from the Cosmic Evolution Early Release Science (CEERS) program\footnote{\url{https://ceers.github.io/}} \citep[][Finkelstein et al. in prep]{finkelstein2022_maisie}, which extends up to $\sim 15$\,arcmin away from EGSY8p7 and $\sim 27$\,arcmin away from \larsonlae, allowing us to search for faint neighbouring systems at large projected separations. Ultimately, this enables us to place constraints on the spatial extent and amplitude of overdensities around the LAEs using a large sample of galaxies, then discuss implications for any ionized bubble that may be present.

This paper is organized as follows. In Section\ \ref{sec:data}, we describe the \jwst\ and ancillary \hst\ imaging used in this work, as well as our method for measuring photometry. We present our selection and sample in Section\ \ref{sec:selection}, then discuss implications for a galaxy overdensity and ionized bubble in Section\ \ref{sec:overdensity}. Finally, we summarize our conclusions in Section\ \ref{sec:summary}. Throughout this work, we adopt a flat $\Lambda$CDM cosmology with $h = 0.7$, $\Omega_\text{m} = 0.3$, and $\Omega_\Lambda = 0.7$. All reported uncertainties correspond to the marginalized 68\,per\,cent credible interval and all logarithms are base-10 unless stated otherwise. Finally, all magnitudes are given the AB system \citep{oke1983} and all distances are physical unless stated otherwise.

\section{Images and Photometry} \label{sec:data}

In this work, we wish to investigate the large-scale neighbourhoods of two extremely UV-luminous ($M_\textsc{uv} \approx -22$) \Lya\ emitters at $z = 8.7$ \citep{zitrin2015, larson2022} in the EGS field. To achieve this, we combine deep multi-band \jwst/NIRCam and \hst/Advanced Camera for Surveys (ACS) imaging over EGS to perform a dropout selection designed to identify sources at redshifts similar to the LAEs (see Section\ \ref{sec:selection} for details). In this section, we describe the imaging data and photometric measurements used for our analysis.
 
The \jwst/NIRCam imaging utilized in this work were taken as part of the CEERS survey in June and December 2022. These NIRCam data consist of imaging in six broadband filters (F115W, F150W, F200W, F277W, F356W, and F444W) and one medium band filter (F410M) over ten independent pointings for a total area of $\sim 85$\,arcmin$^2$. At the redshifts of interest for this work (around $z \sim 8.7$), the NIRCam filters cover a rest-frame wavelength range of $\lambda_\text{rest} \approx 1000-5100$ \AA, probing both the rest-UV (largely redward of the \Lya\ break) and the rest-optical (including the strong emission lines H$\beta$ and \ion{[O}{iii]}$\lambda\lambda$4959,5007 up to $z \sim 8.9$).

We produce co-added mosaics for each NIRCam band following the methods described by \citet{endsley2023_ceers}. We begin from the individual uncalibrated NIRCam exposures ({\tt *\_rate.fits}) and process these exposures with the \textit{JWST} Science Calibration Pipeline,\footnote{\url{https://jwst-pipeline.readthedocs.io/en/latest/index.html}} utilizing the most recent NIRCam photometric calibration reference files (first released as {\tt jwst\textunderscore 0942.pmap} in early October 2022). We align the astrometry of the co-added images for each filter, module, and pointing combination to the \textit{Gaia} astrometric reference frame using the \textsc{tweakreg} package, adopting the \textit{Gaia}-matched mosaic from the \textit{HST}/WFC3 F160W Complete Hubble Archive for Galaxy Evolution (CHArGE) project \citep{kokorev2022} as a reference image. This alignment achieves an rms offset of $\sim 6 - 15$\,mas relative to the \textit{HST} mosaics; see \citet{chen2023} for details on the astrometric calibration. During this process, we also implement a  $1 / f$ noise mitigation algorithm and global background subtraction using the \textsc{sep} package \citep{barbary2016} as described by \citet{endsley2023_ceers}. Finally, all co-added mosaics for each filter are sampled to the same world coordinate system at a resolution of 30\,mas\,pixel$^{-1}$. 

To assist in identifying $z \sim 8.7$ Lyman break galaxies across the CEERS NIRCam footprint, we also utilize optical imaging taken with \textit{HST}/ACS. The EGS field has been observed in the F435W, F606W, and F814W ACS bands as part of the All-Wavelength Extended Groth Strip International Survey (AEGIS; \citealt{davis2007}), the Cosmic Assembly Near-infrared Deep Extragalactic Legacy Survey (CANDELS; \citealt{grogin2011, koekemoer2011}), and the Ultraviolet Imaging of the Cosmic Assembly Near-infrared Deep Extragalactic Legacy Survey Fields (UVCANDELS\footnote{\url{https://archive.stsci.edu/hlsp/uvcandels}}; PI: Teplitz) program. The ACS mosaics used in this work were produced using the \textsc{grizli} software \citep{grizli2022} as part of the CHArGE project. All \textit{HST} mosaics are registered to the \textit{Gaia} astrometric frame and the same world coordinate system with pixel scale of 40\,mas\,pixel$^{-1}$.

The \hst\ and \jwst\ imaging introduced above span more than a factor of three in angular resolution, ranging from a point spread function (PSF) full width at half maximum (FWHM) of $\approx 0.045$\,arcsec in ACS/F435W to $\approx 0.15$\,arcsec in NIRCam/F444W. To ensure robust colour calculations, we must therefore account for variations in the PSF across different images. However, we simultaneously wish to preserve as much signal-to-noise as possible, particularly in the shorter wavelength filters that probe the \Lya\ break. We thus homogenize the PSFs of all the ACS and NIRCam short wavelength (SW; F115W, F150W, F200W) mosaics to that of ACS/F814W, while the NIRCam long wavelength (LW; F277W, F356W, F410M, F444W) mosaics are homogenized to the PSF of WFC3/F160W to enable future studies using \textit{HST}/WFC3.

\begin{figure}
    \centering
    \includegraphics[width=\columnwidth]{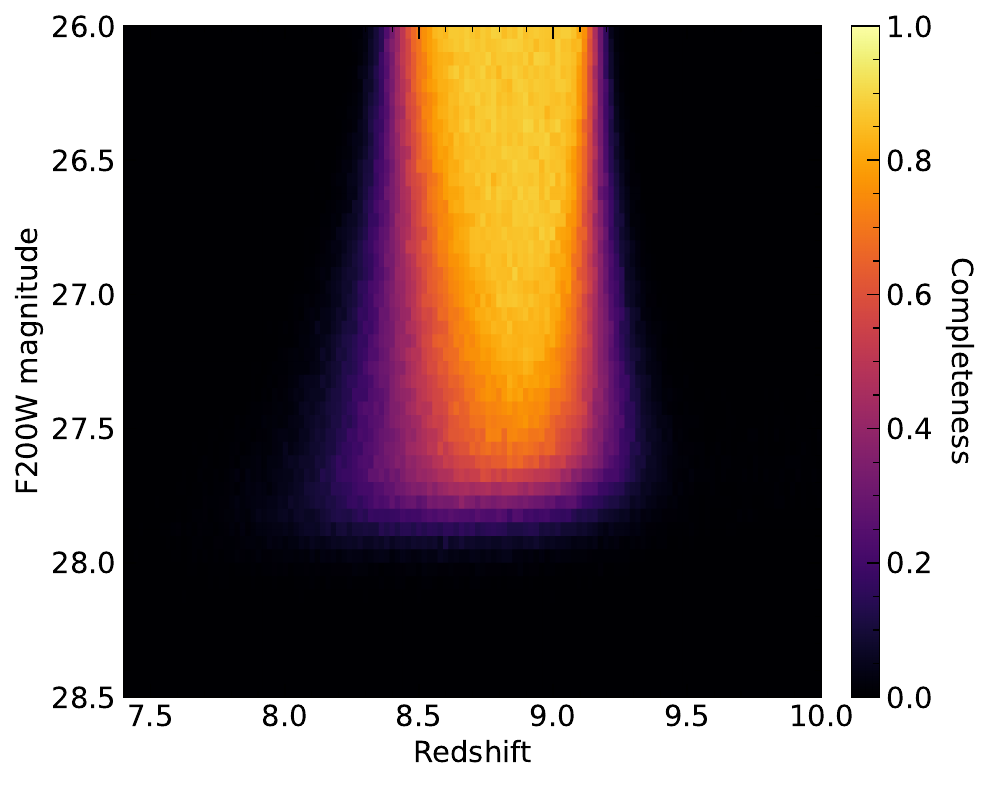}
    \caption{Simulated completeness for our colour and S/N selection criteria designed to identify bright objects at redshifts close to $z \sim 8.7$ (see Section\ \ref{subsec:selection}). The selection generally identifies $\text{F200W} = 27$ objects at redshifts of $z \sim 8.4 - 9.1$ with $\gtrsim 50$\,per\,cent completeness. Applying this selection to the CEERS NIRCam imaging, we identify 27 galaxy candidates, one of which is EGSY8p7.}
    \label{fig:completeness}
\end{figure}

We identify sources across the CEERS NIRCam footprint by running \textsc{Source Extractor} on an inverse variance weighted stack of the PSF-homogenized F150W and F200W mosaics. The variance images are computed by running \textsc{sep} on each mosaic and thus incorporate both background and detector noise. We calculate photometry following procedures previously used for \hst-based analyses of $z\gtrsim6$ galaxies \citep[e.g.][]{bouwens2015, bouwens2021, finkelstein2015, finkelstein2022, endsley2021_OIII}. Our methods are described in full by \citet{endsley2023_ceers}, but we provide a brief overview here. First, we use the background-subtracted mosaics to compute photometry in elliptical \citet{kron1980} apertures following procedures similar to those used for previous analyses of $z \gtrsim 6$ galaxies \citep[e.g.][]{bouwens2015, bouwens2021, bouwens2023_jems, finkelstein2015, finkelstein2022, finkelstein2023_ceers, endsley2021_OIII}, adopting aperture sizes set by a Kron parameter of $k = 1.2$ to maximize signal-to-noise \citep{finkelstein2022}. Photometric uncertainties for each object in the catalog are calculated as the standard deviation of flux measured in apertures of the same size as the aperture used for the object, which are randomly placed in nearby object-free regions as determined from a \textsc{Source Extractor} segmentation map specific to each band. Finally, we perform aperture corrections in two stages to obtain total fluxes (see section 2.3 of \citealt{finkelstein2022} for details). First, we multiply all ACS and SW fluxes and their errors by the ratio of F200W flux measured in apertures with $k=2.5$ versus $k=1.2$ from the PSF-homogenized F200W mosaic for ACS and the NIRCam SW filters, and an inverse variance weighted stack of the LW mosaics for the LW photometry. Second, the photometry is corrected for flux outside the $k = 2.5$ aperture using the measured PSFs of ACS/F814W and WFC3/F160W from the mosaics, normalizing the encircled energy distribution at large radii ($> 1$\,arcsec) based on reported values in the instrument manuals. To account for the different PSFs of the ACS+SW and LW images, we multiply the aperture size of the ACS+SW bands by a factor of 1.5 to obtain the aperture for the LW photometry, where the factor of 1.5 reflects the typical size ratio of twelve UV-bright $z\sim6-8$ EGS galaxy candidates \citep{chen2023} measured from the PSF-homogenized F200W and F444W mosaics.

Though the NIRCam images have a relatively high angular resolution, it is nevertheless possible for flux from neighbouring objects to contaminate our photometric apertures. To mitigate the possibility of such contaminated photometry, we have developed a neighbour subtraction algorithm heavily based on techniques developed for \textit{Spitzer}/Infrared Array Camera (IRAC) deconfusion \citep[e.g.][]{labbe2010, labbe2013, bouwens2015, endsley2021_OIII} and described in detail by \citet{endsley2023_ceers}. In brief, given an object in the catalog, we fit S{\'e}rsic profiles to neighbours identified in the \textsc{Source Extractor} catalog using the non-PSF matched mosaics, accounting for the PSF of each filter during the fit. Using the best-fitting S\'ersic profile parameters, we then subtract the neighbouring objects in the PSF-homogenized images, accounting for the changes in the PSF. We apply this algorithm to sources in our colour selected sample (selected as described in Section\ \ref{subsec:selection}) that are contaminated by flux from a neighbouring object within the $k = 2.5$ Kron aperture as determined by a visual inspection.

\begin{figure*}
    \centering
    \includegraphics[width=\textwidth]{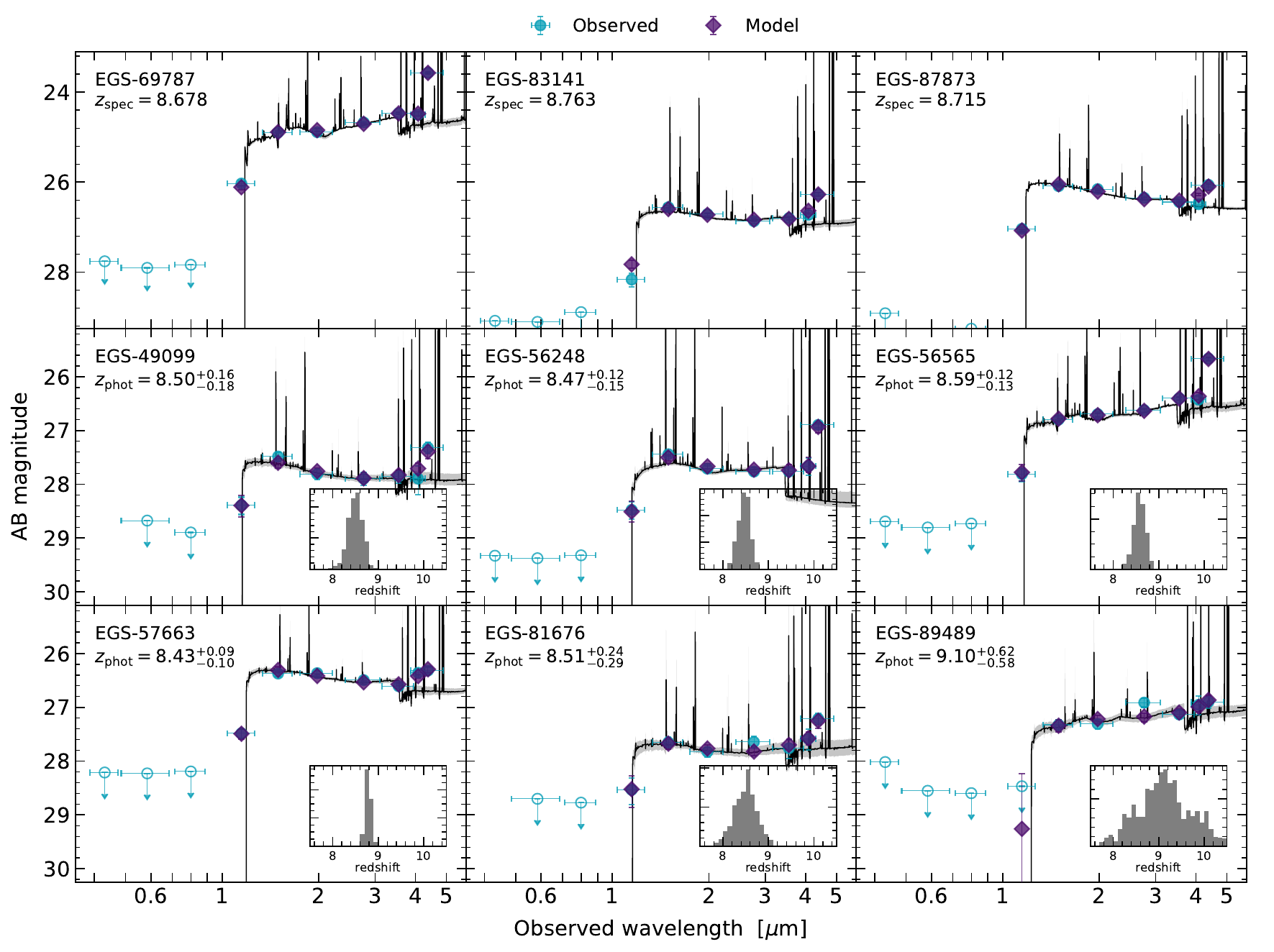}
    \caption{The observed and model SEDs of a subset of our sample: three objects with spectroscopic redshifts (top row), and the six objects closest to EGSY8p7 (all of which are within 3\,arcmin of EGSY8p7; center and bottom rows). The observed photometry is shown as teal circles, with open circles denoting $2\sigma$ upper limits, and the model photometry is shown as purple diamonds. For the objects without a spectroscopic redshift, we also show the photometric redshift posteriors that result from the models restricted to photometric redshifts of $z_\text{phot} = 4 - 12$ described in Section\ \ref{subsec:sample}. We highlight that most of these objects have red observed $\text{F410M} - \text{F444W}$ colors ($\text{F410M} - \text{F444W} \gtrsim 0.5$), which is likely due to strong \ion{[O}{iii]}+\ion{H}{$\beta$} emission boosting the F444W flux; a 0.5\,mag F444W excess approximately corresponds to an EW of $\sim 650$\,\AA\ at $z = 8.7$.}
    \label{fig:seds}
\end{figure*}

\begin{table*}
\renewcommand{\arraystretch}{1.5}
\centering
\caption{Observational and physical properties of the candidates identified with our colour and S/N selection criteria (Section\ \ref{subsec:selection}) ordered by increasing angular separation from EGSY8p7. We report the observed F200W magnitudes and observed rest-UV continuum slopes. We also report properties inferred from our \beagle\ models: the photometric redshifts (or systemic spectroscopic redshift without errors, if available), stellar masses, sSFRs, and intrinsic ionizing photon production efficiencies before processing through dust and gas. For these SED models, we have assumed that these objects lie at $z \geq 4$ to ensure that our inferences are representative of the physical properties these systems would have at high redshift rather than the properties these objects would have at lower redshifts; see Section\ \ref{subsec:sample}.}
\label{tab:sample_properties}
\begin{tabular}{c|c|c|c|c|c|c|c|c|c|c} \hline
    \multirow{2}{*}{ID} & \multirow{2}{*}{RA} & \multirow{2}{*}{Dec} & \multirow{2}{*}{Redshift} & \multirow{2}{*}{$m_\text{F200W}$} & \multirow{2}{*}{$\beta$} & \multirow{2}{*}{$\log(M_* / M_\odot)$} & \multirow{2}{*}{Age [Myr]} & sSFR & $\log(\xi_{\text{ion}}^{*})$ & \multirow{2}{*}{\zsys\ ref.} \\
    & & & & & & & & [Gyr$^{-1}$] & [Hz erg$^{-1}$] & \\ \hline\hline
    69787* & 14:20:08.49 & +52:53:26.40 & 8.678 & $24.9_{-0.0}^{+0.0}$ & $-1.7_{-0.0}^{+0.0}$ & $9.0_{-0.1}^{+0.1}$ & $6_{-1}^{+1}$ & $173_{-21}^{+20}$ &  $25.74_{-0.02}^{+0.02}$ & 1, 2, 3, 4 \\
    57663 & 14:20:02.08 & +52:52:05.90 & $8.78_{-0.06}^{+0.06}$ & $26.2_{-0.0}^{+0.0}$ & $-1.8_{-0.1}^{+0.1}$ & $8.5_{-0.5}^{+0.4}$ & $27_{-18}^{+59}$ & $37_{-25}^{+77}$ &  $25.68_{-0.13}^{+0.17}$ & --- \\
    81676 & 14:20:19.11 & +52:54:34.44 & $8.50_{-0.25}^{+0.22}$ & $27.8_{-0.1}^{+0.1}$ & $-2.0_{-0.2}^{+0.2}$ & $8.0_{-0.7}^{+0.6}$ & $28_{-23}^{+136}$ & $36_{-30}^{+179}$ &  $25.59_{-0.12}^{+0.31}$ & --- \\
    56565 & 14:19:57.50 & +52:51:59.61 & $8.59_{-0.13}^{+0.12}$ & $26.7_{-0.0}^{+0.0}$ & $-1.7_{-0.1}^{+0.1}$ & $8.3_{-0.3}^{+0.2}$ & $12_{-6}^{+10}$ & $84_{-38}^{+83}$ &  $25.66_{-0.06}^{+0.12}$ & --- \\
    56248 & 14:19:56.21 & +52:51:57.91 & $8.47_{-0.15}^{+0.12}$ & $27.7_{-0.1}^{+0.1}$ & $-2.6_{-0.1}^{+0.1}$ & $7.5_{-0.1}^{+0.1}$ & $2_{-0}^{+1}$ & $601_{-205}^{+245}$ &  $26.04_{-0.03}^{+0.02}$ & --- \\
    49099 & 14:19:59.13 & +52:51:14.96 & $8.50_{-0.17}^{+0.16}$ & $27.8_{-0.1}^{+0.1}$ & $-2.7_{-0.2}^{+0.2}$ & $7.9_{-0.4}^{+0.5}$ & $33_{-22}^{+107}$ & $31_{-24}^{+65}$ &  $25.59_{-0.10}^{+0.19}$ & --- \\
    89489 & 14:20:16.84 & +52:55:45.48 & $9.10_{-0.58}^{+0.62}$ & $27.3_{-0.1}^{+0.1}$ & $-1.3_{-0.2}^{+0.2}$ & $8.5_{-0.6}^{+0.5}$ & $35_{-27}^{+97}$ & $28_{-21}^{+93}$ &  $25.62_{-0.12}^{+0.21}$ & --- \\
    83130 & 14:19:48.14 & +52:54:42.39 & $8.52_{-0.16}^{+0.17}$ & $26.9_{-0.1}^{+0.1}$ & $-2.6_{-0.3}^{+0.3}$ & $8.4_{-0.3}^{+0.3}$ & $45_{-25}^{+53}$ & $22_{-12}^{+27}$ &  $25.61_{-0.07}^{+0.10}$ & --- \\
    89540 & 14:19:52.49 & +52:55:46.75 & $8.84_{-0.18}^{+0.10}$ & $26.6_{-0.1}^{+0.1}$ & $-2.1_{-0.2}^{+0.2}$ & $7.9_{-0.1}^{+0.6}$ & $6_{-4}^{+26}$ & $166_{-135}^{+480}$ &  $25.85_{-0.28}^{+0.19}$ & --- \\
    87873 & 14:19:52.21 & +52:55:58.65 & 8.715 & $26.2_{-0.0}^{+0.0}$ & $-2.3_{-0.1}^{+0.1}$ & $8.5_{-0.1}^{+0.2}$ & $30_{-8}^{+15}$ & $33_{-11}^{+12}$ &  $25.64_{-0.04}^{+0.04}$ & 2, 5 \\
    83141 & 14:19:45.27 & +52:54:42.32 & 8.763 & $26.7_{-0.0}^{+0.0}$ & $-2.5_{-0.1}^{+0.1}$ & $7.8_{-0.2}^{+0.2}$ & $7_{-4}^{+6}$ & $147_{-69}^{+190}$ &  $25.84_{-0.11}^{+0.16}$ & 6 \\
    57722 & 14:19:41.84 & +52:52:06.94 & $8.80_{-0.34}^{+0.34}$ & $28.0_{-0.1}^{+0.2}$ & $-2.5_{-0.4}^{+0.4}$ & $7.8_{-0.5}^{+0.6}$ & $26_{-20}^{+118}$ & $38_{-31}^{+138}$ &  $25.58_{-0.11}^{+0.22}$ & --- \\
    44155 & 14:19:46.52 & +52:50:39.22 & $8.43_{-0.13}^{+0.18}$ & $26.7_{-0.1}^{+0.1}$ & $-2.4_{-0.2}^{+0.2}$ & $8.2_{-0.6}^{+0.4}$ & $29_{-27}^{+54}$ & $35_{-23}^{+479}$ &  $25.65_{-0.10}^{+0.40}$ & --- \\
    130899 & 14:19:48.02 & +52:56:57.38 & $8.21_{-0.21}^{+0.27}$ & $27.9_{-0.1}^{+0.2}$ & $-2.9_{-0.4}^{+0.4}$ & $7.6_{-0.3}^{+0.5}$ & $13_{-10}^{+36}$ & $77_{-56}^{+218}$ &  $25.65_{-0.12}^{+0.26}$ & --- \\
    126832 & 14:19:47.34 & +52:57:23.96 & $8.32_{-0.19}^{+0.18}$ & $27.5_{-0.1}^{+0.1}$ & $-2.3_{-0.3}^{+0.3}$ & $8.0_{-0.4}^{+0.5}$ & $35_{-23}^{+92}$ & $29_{-21}^{+56}$ &  $25.62_{-0.11}^{+0.16}$ & --- \\
    91290 & 14:19:34.01 & +52:55:35.21 & $8.71_{-0.70}^{+0.35}$ & $27.5_{-0.1}^{+0.1}$ & $-2.1_{-0.3}^{+0.3}$ & $8.8_{-0.8}^{+0.2}$ & $191_{-174}^{+241}$ & $5_{-3}^{+54}$ &  $25.47_{-0.02}^{+0.18}$ & --- \\
    118482 & 14:19:48.43 & +52:58:19.59 & $8.19_{-0.19}^{+0.22}$ & $27.5_{-0.1}^{+0.2}$ & $-2.1_{-0.3}^{+0.3}$ & $7.7_{-0.3}^{+0.6}$ & $12_{-10}^{+51}$ & $83_{-67}^{+432}$ &  $25.75_{-0.22}^{+0.30}$ & --- \\
    118130 & 14:20:28.81 & +52:58:21.22 & $8.58_{-0.16}^{+0.15}$ & $27.5_{-0.1}^{+0.1}$ & $-2.3_{-0.2}^{+0.2}$ & $7.9_{-0.5}^{+0.5}$ & $24_{-18}^{+73}$ & $42_{-32}^{+140}$ &  $25.67_{-0.15}^{+0.24}$ & --- \\
    64770 & 14:19:30.27 & +52:52:50.99 & $7.69_{-0.01}^{+0.01}$ & $27.4_{-0.0}^{+0.0}$ & $-1.5_{-0.1}^{+0.1}$ & $7.9_{-0.1}^{+0.1}$ & $4_{-1}^{+1}$ & $268_{-65}^{+72}$ &  $25.84_{-0.04}^{+0.03}$ & --- \\
    110201 & 14:20:02.81 & +52:59:17.91 & 8.876 & $26.6_{-0.0}^{+0.0}$ & $-2.6_{-0.1}^{+0.1}$ & $8.4_{-0.2}^{+0.2}$ & $34_{-14}^{+29}$ & $29_{-13}^{+20}$ &  $25.66_{-0.08}^{+0.08}$ & 5, 7 \\
    109389 & 14:19:58.65 & +52:59:21.77 & 8.807 & $27.1_{-0.1}^{+0.1}$ & $-2.4_{-0.1}^{+0.1}$ & $8.2_{-0.3}^{+0.4}$ & $32_{-19}^{+71}$ & $31_{-22}^{+49}$ &  $25.63_{-0.11}^{+0.14}$ & 7 \\
    116241 & 14:20:37.47 & +52:58:36.06 & $8.72_{-0.22}^{+0.21}$ & $27.6_{-0.1}^{+0.1}$ & $-2.0_{-0.3}^{+0.3}$ & $8.2_{-0.7}^{+0.6}$ & $38_{-32}^{+163}$ & $26_{-21}^{+145}$ &  $25.57_{-0.10}^{+0.29}$ & --- \\
    70289 & 14:19:22.74 & +52:53:31.60 & $8.62_{-0.12}^{+0.12}$ & $26.9_{-0.1}^{+0.1}$ & $-1.8_{-0.1}^{+0.1}$ & $8.2_{-0.1}^{+0.1}$ & $4_{-1}^{+1}$ & $228_{-53}^{+67}$ &  $25.83_{-0.05}^{+0.06}$ & --- \\
    101537 & 14:20:05.90 & +53:00:39.25 & $8.56_{-0.21}^{+0.18}$ & $27.5_{-0.1}^{+0.1}$ & $-1.7_{-0.2}^{+0.2}$ & $7.6_{-0.1}^{+0.6}$ & $2_{-1}^{+28}$ & $456_{-424}^{+311}$ &  $26.04_{-0.40}^{+0.04}$ & --- \\
    15298 & 14:19:20.26 & +52:46:53.92 & $8.44_{-0.15}^{+0.18}$ & $27.7_{-0.1}^{+0.1}$ & $-2.2_{-0.3}^{+0.3}$ & $7.9_{-0.4}^{+0.5}$ & $27_{-18}^{+73}$ & $37_{-27}^{+69}$ &  $25.62_{-0.11}^{+0.18}$ & --- \\
    35943 & 14:18:58.14 & +52:49:43.77 & $8.68_{-0.16}^{+0.14}$ & $26.8_{-0.1}^{+0.1}$ & $-1.8_{-0.2}^{+0.2}$ & $8.6_{-0.5}^{+0.3}$ & $50_{-36}^{+88}$ & $20_{-13}^{+51}$ &  $25.60_{-0.09}^{+0.17}$ & --- \\
    2992 & 14:19:05.48 & +52:44:27.11 & $8.84_{-0.14}^{+0.27}$ & $26.7_{-0.1}^{+0.1}$ & $-2.0_{-0.2}^{+0.2}$ & $7.9_{-0.1}^{+0.6}$ & $5_{-3}^{+30}$ & $215_{-187}^{+472}$ &  $25.93_{-0.35}^{+0.15}$ & --- \\ \hline
\end{tabular} \\
\raggedright$^*$EGSY8p7 \citep{zitrin2015} \\
\raggedright References for \zsys: [1] \citet{isobe2023}; [2] \citet{tang2023}; [3] \citet{sanders2023}; [4] \citet{larson2023}; [5] \citet{nakajima2023}; [6] \citet{arrabalharo2023a}; [7] \citet{fujimoto2023}
\end{table*}

\begin{table*}
\renewcommand{\arraystretch}{1.5}
\centering
\caption{Coordinates, angular separations, and physical separations from EGSY8p7 for our sample, ordered by increasing angular separation from EGSY8p7. If a systemic spectroscopic redshift for a given object is available (denoted by a superscript dagger next to the object ID), we adopt the systemic redshift to calculate the physical separation from EGSY8p7. Otherwise, we adopt the median photometric redshift of the object inferred by \beagle.}
\label{tab:sample_locations}
\begin{tabular}{c|c|c|c|c} \hline
    \multirow{2}{*}{ID} & \multirow{2}{*}{RA} & \multirow{2}{*}{Dec} & \multicolumn{2}{c}{Distance from EGSY8p7} \\
    & & & [arcmin] & [pMpc] \\ \hline\hline
    69787* & 14:20:08.49 & +52:53:26.40 & --- & --- \\
    57663 & 14:20:02.08 & +52:52:05.90 & 1.7 & 2.7 \\
    81676 & 14:20:19.11 & +52:54:34.44 & 2.0 & 4.9 \\
    56565 & 14:19:57.50 & +52:51:59.61 & 2.2 & 2.5 \\
    56248 & 14:19:56.21 & +52:51:57.91 & 2.4 & 5.8 \\
    49099 & 14:19:59.13 & +52:51:14.96 & 2.6 & 5.0 \\
    89489 & 14:20:16.84 & +52:55:45.48 & 2.6 & 10.8 \\
    83130 & 14:19:48.14 & +52:54:42.39 & 3.3 & 4.3 \\
    89540 & 14:19:52.49 & +52:55:46.75 & 3.4 & 4.4 \\
    87873$^{\dagger}$ & 14:19:52.21 & +52:55:58.65 & 3.5 & 1.4 \\
    83141$^{\dagger}$ & 14:19:45.27 & +52:54:42.32 & 3.7 & 2.5 \\
    57722 & 14:19:41.84 & +52:52:06.94 & 4.2 & 3.3 \\
    44155 & 14:19:46.52 & +52:50:39.22 & 4.3 & 7.1 \\
    130899 & 14:19:48.02 & +52:56:57.38 & 4.7 & 13.3 \\
    126832 & 14:19:47.34 & +52:57:23.96 & 5.1 & 10.3 \\
    91290 & 14:19:34.01 & +52:55:35.21 & 5.6 & 1.8 \\
    118482 & 14:19:48.43 & +52:58:19.59 & 5.7 & 13.9 \\
    118130 & 14:20:28.81 & +52:58:21.22 & 5.8 & 3.2 \\
    64770 & 14:19:30.27 & +52:52:50.99 & 5.8 & 30.1 \\
    110201$^{\dagger}$ & 14:20:02.81 & +52:59:17.91 & 5.9 & 5.4 \\
    109389$^{\dagger}$ & 14:19:58.65 & +52:59:21.77 & 6.1 & 3.8 \\
    116241 & 14:20:37.47 & +52:58:36.06 & 6.8 & 2.2 \\
    70289 & 14:19:22.74 & +52:53:31.60 & 6.9 & 2.5 \\
    101537 & 14:20:05.90 & +53:00:39.25 & 7.2 & 3.9 \\
    15298 & 14:19:20.26 & +52:46:53.92 & 9.8 & 7.2 \\
    35943 & 14:18:58.14 & +52:49:43.77 & 11.3 & 3.1 \\
    2992 & 14:19:05.48 & +52:44:27.11 & 13.1 & 5.6 \\ \hline
\end{tabular} \\
\raggedright$^*$EGSY8p7 \citep{zitrin2015} \\
\raggedright$^\dagger$Object has a spectroscopic redshift that was used to calculate physical separation. See references in Table\ \ref{tab:sample_properties}.
\end{table*}

\section{Sample selection and properties} \label{sec:selection}

We now wish to build a sample designed to assess whether there is evidence of any large-scale galaxy overdensities surrounding the two extremely UV-luminous ($M_\textsc{uv} \approx -22$) \Lya\ emitters at $z = 8.7$ in the EGS field. These LAEs lie at nearly the same redshift \citep[EGSY8p7 at $\zsys = 8.678$ and \larsonlae\ at $\zsys = 8.610$;][]{tang2023, larson2023} and are separated by only 4\,pMpc, suggesting they may be part of the same overdensity powering a very large ($R \gtrsim 2$\,pMpc) ionized bubble \citep{larson2022}.

To investigate the possibility of a large-scale $z \sim 8.7$ overdensity within EGS, we identify Lyman break galaxies at similar photometric redshifts over the entire CEERS NIRCam imaging footprint. This footprint extends to approximately $\sim 15$ ($\sim 27$) arcmin away from EGSY8p7 (\larsonlae), corresponding to maximum separations of $\sim 4$ (7.5) pMpc in projection at $z = 8.7$. As the number of $z \sim 9$ galaxy candidates in EGS that have thus far been targeted with sensitive rest-optical spectroscopy is limited, in this work we aim to quantify the overdensity using photometrically selected Lyman break galaxies (as in, e.g. \citealt{castellano2016, endsley2021_LyA, leonova2022}). Consequently, we focus on angular separations, acknowledging that the full 3D separations can be larger given the redshift uncertainties associated with broadband dropout selections (e.g. a redshift separation of $\Delta z = 0.1$ corresponds to a line-of-sight separation of $\sim 2.6$\,pMpc at $z = 8.7$).

\subsection{Selection criteria} \label{subsec:selection}

At $z \sim 8.7$, the \Lya\ break falls in the NIRCam F115W filter ($\sim 40$\,per\,cent of the integrated transmission contains flux redward of the break). We therefore expect sources at redshifts similar to that of the LAEs to appear as complete ACS/F814W dropouts and partial NIRCam/F115W dropouts. Accordingly, we adopt the following selection criteria to identify these objects:
\begin{enumerate}
    \item $\text{F200W} < 28$ and $\text{F200W S/N} > 7$
    \item $\text{S/N} > 3$ in at least two LW filters
    \item $\text{S/N} < 2$ in F435W, F606W, and F814W
    \item $\chi^2_\text{opt} < 5$ (see footnote\footnote{We calculate the `optical $\chi^2$' ($\chi^2_\text{opt}$) criteria previously used for \textit{HST}-based studies of $z > 7$ galaxies \citep[e.g.][]{bouwens2015} as $\chi^2_\text{opt} = \sum_{i}\text{sgn}(f_i) \, (f_i / \sigma_i)^2$, where $f_i$ and $\sigma_i$ are the observed flux and error in filter $i$ for a given object, $\text{sgn}(f_i) = 1$ if $f_i > 0$ and $\text{sgn}(f_i) = -1$ if $f_i < 0$, and the sum runs over the three optical ACS filters.})
    \item $\text{F814W} - \text{F150W} > 1.7$
    \item $\text{F115W} - \text{F150W} > 0.6$ and $\text{F115W} - \text{F150W} < 1.7$
    \item $\text{F150W} - \text{F277W} < 0.6$
    \item $\text{F115W} - \text{F150W} > 1.5 \times \left(\text{F150W} - \text{F277W}\right) + 0.6$.
\end{enumerate}
We first require our sample to have apparent magnitudes of $\text{F200W} < 28$, $\text{S / N} > 7$ in F200W, and $\text{S / N} > 3$ in at least two LW NIRCam filters to ensure the selected sources are real (criteria (i) and (ii)). Next, criteria (iii), (iv), and (v) are designed to ensure the presence of a complete \Lya\ break between ACS/F814W and NIRCam/F150W and effectively no flux blueward of the break, as expected at $z > 8$ from the very high Lyman series and Lyman continuum opacity through the partially neutral IGM \citep{inoue2014}. We note that fluxes in the F814W and F115W dropout filters are set to their $1\sigma$ upper limits if detected at $< 1\sigma$ significance. Criterion (vi) is designed to identify partial F115W dropouts while selecting against a complete \Lya\ break in F115W to place an upper limit on our redshift range. Finally, criteria (vii) and (viii) are designed to remove candidates with extremely red spectral energy distributions (SEDs) in order to allow us to distinguish $z > 8$ sources, which are typically blue \citep[e.g.][]{bouwens2014, bhatawdekar2021, topping2022, cullen2023}, from intrinsically red galaxies at lower redshifts. We note that there is no ACS/F435W imaging for $\sim 15$\,arcmin$^2$ of the total $\sim 85$\,arcmin$^2$ NIRCam imaging footprint. For these areas, we adjust criterion (iii) to require $\text{S/N} < 2$ only in F606W and F814W and adjust criterion (iv) to $\chi^2_\text{opt} < 3.5$, as we only calculate $\chi^2_\text{opt}$ using two filters.

In general, we expect this selection to efficiently identify moderately faint objects at redshifts of $z \sim 8.4 - 9.1$. We estimate our selection completeness by generating mock power law SEDs ($f_\lambda \propto \lambda^{\beta}$) with rest-UV continuum slope of $\beta = -2$. We normalize these SEDs to a range of F200W magnitudes ($\text{F200W} = 24 - 30$, $\text{d}m = 0.05$) at a variety of redshifts ($z = 5 - 12$, $\text{d}z = 0.02$), applying the IGM attenuation model of \citet{inoue2014}. For our fiducial completeness models, we assume a \Lya\ EW of 0\,\AA, but explore the impact of $\text{EW}_{\Lya} > 0$\,\AA\ below. We then convolve the normalized SEDs with the transmission functions of our ten ACS and NIRCam filters to obtain mock observations, perturb the `observed' fluxes appropriately for the depth of each filter as measured from real sources, and apply the selection criteria described above. Using these models, we estimate that our colour selection is $\geq 50$\,per\,cent ($\geq 80$\,per\,cent) complete from $z \sim 8.4 - 9.1$ ($z \sim 8.5 - 9.1$) at $\text{F200W} = 27$ with a maximum efficiency of $\sim 90$\,per\,cent due to upscattering in the optical ACS bands and our requirements of ACS nondetections and $\chi^2_\text{opt} < 5$. At fainter magnitudes, $\text{F200W} \sim 27.5$, our completeness function broadens in redshift as objects at slightly lower ($z \sim 8.1 - 8.4$) and higher ($z \sim 9.1 - 9.4$) redshifts scatter into our selection with $\lesssim 10$\,per\,cent completeness.

If the galaxies we wish to select reside in an overdensity, and associated large ionized bubble, we may expect that they have reasonably strong \Lya\ emission. Thus, we wish to ensure that our selection would identify such objects at redshifts similar to those of the two known LAEs. At $z \sim 8.7$, \Lya\ transmits through $\text{F115W}$, resulting in bluer $\text{F115W} - \text{F150W}$ colours at fixed redshift if a galaxy has strong \Lya\ emission, ultimately shifting our selection to higher redshifts for LAEs. To quantify this effect, we model our selection completeness adopting \Lya\ EWs ranging from 5\,\AA\ to 50\,\AA. For \Lya\ $\text{EW}_\Lya = 10$\,\AA\ \citep[approximately the median value for $M_\textsc{uv} \lesssim -20.5$ galaxies at $z \sim 7$ found by][]{endsley2021_LyA}, our selection identifies moderately faint ($\text{F200W} < 28$) objects at redshifts of $z \sim 8.5 - 9.3$. For stronger \Lya\ ($\text{EW}_\Lya = 30$\,\AA), the selection identifies candidates at redshifts of $z \sim 8.6 - 9.4$, still near the redshift of the LAEs. In particular, we highlight that our selection recovers EGSY8p7 itself (EGS-69787 in our catalog), which is at a systemic redshift of $\zsys = 8.678$ \citep{tang2023, larson2023} with $\text{EW}_\Lya \approx 10 - 30$\,\AA\ (lower bound from \textit{JWST}/NIRSpec observations, \citealt{tang2023, larson2023}; upper bound from ground-based Keck/MOSFIRE observations, which is impacted by an OH sky line, \citealt{zitrin2015}). We would also expect our selection to identify \larsonlae\ at $\zsys = 8.610$ \citep{tang2023} with $\text{EW}_\Lya \approx 5$\,\AA\ \citep{larson2022}, though this source falls outside the CEERS NIRCam footprint.

To further ensure that our conclusions do not depend strongly on the possible lack of the strongest LAEs ($\text{EW}_\Lya \gtrsim 75$\,\AA) at $z \sim 8.7$ identified by our primary selection, we design a secondary selection to identify these objects, described in Appendix\ \ref{appendix:lae_selection}. Using this selection, we identify 13 additional candidates over the field, one of which lies 0.7\,arcmin away from EGSY8p7. We highlight that, in general, our conclusions do not significantly change if we supplement our fiducial sample with these potential strong LAEs at $z \sim 8.7$.

\bigskip

After applying our fiducial selection criteria, we conduct a visual inspection of the objects that are identified. We remove sources that (1) have photometry that is likely contaminated by a nearby bright source or artifact (primarily diffraction spikes or `snowballs' caused by a significant cosmic ray event), (2) appear to have by-eye detections in any of the three optical ACS bands \citep[which are expected to be essentially completely attenuated by the IGM at our redshifts of interest;][]{inoue2014}, or (3) appear to have unreliable noise measurements (often due to being near a detector edge). We additionally remove one object (EGS-25508) with a pattern of photometric excesses in F150W, F200W, and F277W, followed by a drop in flux in the longer wavelength filters that is highly suggestive of emission lines at $z \sim 2$ boosting the $1.5 - 3$\,$\mu$m filters, and for which the SED models described in Section\ \ref{subsec:photozs} find no acceptable solutions at $z \geq 6$.

After visual inspection, we obtain a final colour-selected sample of 27 dropout candidates (one of which is EGSY8p7), which we use as our fiducial sample for the remainder of this work. However, for the interested reader, we summarize the objects that we removed during our visual inspection in Appendix\ \ref{appendix:visual_inspection}. For each of the 27 objects in our fiducial sample, we perform a second visual inspection to determine whether the $k = 2.5$ Kron apertures used for photometry are contaminated by flux from a neighbouring object, and apply the neighbour subtraction algorithm described in Section\ \ref{sec:data} when necessary to remove contaminating flux. Otherwise, we use the non-neighbour subtracted photometry. During this visual inspection, we also identify two candidates (EGS-44155 and EGS-70289) that consist of multiple components. For these two systems, we perform photometry (following the methods of \citealt{chen2023} and \citealt{endsley2023_ceers}) with manual elliptical apertures to ensure that our photometry reflects the integrated flux from all components.

\bigskip

Before presenting our sample in detail, we investigate whether $z \sim 9$ candidates previously identified with \textit{HST} imaging in EGS satisfy our selection criteria. \citet{bouwens2021} reported four objects with photometric redshifts of $z_\text{phot} = 8.2 - 9.2$ (EGSY-9597051195, EGS910-3, EGS910-10, and EGS910-15) that fall within the CEERS imaging footprint, three of which we recover \citep[one of which is EGSY8p7, and the other two of which are now confirmed to lie at $z_\text{spec} = 8.7$;][]{tang2023, arrabalharo2023a}. We find that the fourth, EGSY-9597051195, has only a weak break between F115W and F150W ($\text{F115W} - \text{F150W} = 0.1$) and is consequently best fit by a photometric redshift of $z_\text{phot} = 7.2$, lower than we consider in this work. \citet{finkelstein2022} also reported four \textit{HST}-selected objects at $z_\text{phot} = 8.2 - 9.2$ that lie within the CEERS imaging (EGS\textunderscore z910\textunderscore 6811, EGS\textunderscore z910\textunderscore 68560, EGS\textunderscore z910\textunderscore 26890, and EGS\textunderscore z910\textunderscore 40898), all but one of which we select (noting that EGS\textunderscore z910\textunderscore 6811 is EGSY8p7 and EGS\textunderscore z910\textunderscore 26890 is the same object as EGS910-15 identified by \citealt{bouwens2021}). EGS\textunderscore z910\textunderscore 68560 is not detected within 1\,arcsec of its expected position in the NIRCam imaging; given its reported F160W magnitude of $m_\textsc{ab} = 25.8$ reported by \citet{finkelstein2022} and the depth of the \textit{JWST} imaging ($5\sigma$ depth of $\sim 28.3$\,mag in F150W), this suggests that this object may be a transient source.

Additionally, \citet{leonova2022} conducted a targeted \textit{HST}/WFC3 search for galaxy candidates that lie very close to extremely luminous, high redshift galaxies in the EGS field. They identified eight objects, including EGSY8p7 and EGS910-3, and our selection recovers four (EGSY8p7, EGS910-3, EGSY-66990, and EGSY-69674). One of the candidates that we do not identify falls in a detector gap in the SW NIRCam imaging of CEERS \citep[EGSY-27727 in][]{leonova2022}. Another (EGSY-56680, with reported F160W magnitude of $m_\textsc{ab} = 27.2$) is not detected in the CEERS imaging which, as for EGS\textunderscore z910\textunderscore 68560, may imply that this object could be a transient source. Finally, the remaining two objects we do not select (EGSY-6887 and EGSY-12580) both have bluer $\text{F115W} - \text{F150W}$ colours than we allow during our colour selection ($\text{F115W} - \text{F150W} = 0.4$ and $-0.2$, respectively) as well as evidence of F410M excesses consistent with strong rest-optical emission lines transmitting through F410M at slightly lower redshifts than we target in this work ($z \sim 7 - 7.5$).

We emphasize that the identification of objects by \textit{HST} that are likely at slightly lower redshifts than we target is unsurprising. Photometric redshift uncertainties with \textit{HST} photometry can be large (e.g. \citealt{leonova2022} reported median upper uncertainties of $\Delta z \approx 0.2$ and median lower uncertainties of $\Delta z \approx 0.5$). Furthermore, while objects down to $z \sim 7.5$ drop out at least partially in WFC3/F105W, objects at $z \sim 7.5 - 8$ do not significantly drop out in NIRCam/F115W, allowing us to cleanly remove these slightly lower redshift objects from our sample. Overall, the availability of the NIRCam/F115W filter significantly improves our ability to distinguish between objects at $z \sim 7.5 - 8$ and $z \sim 9$. Moreover, overlapping \textit{JWST}/NIRCam broad and medium bands then enable us to constrain the photometric redshifts of candidates via strong emission lines even more precisely.

We further highlight that our selection recovers \textit{all} of the \textit{HST}-selected galaxies in the CEERS footprint that have now been confirmed to lie at $z_\text{spec} = 8.7$ \citep{tang2023, sanders2023, fujimoto2023, larson2023, arrabalharo2023a, arrabalharo2023b, harikane2023_specz}. All of these objects (EGS-69787, EGS-83141, and EGS-87873 in our catalog) have evidence of F444W excesses likely caused by \ion{[O}{iii]}+\ion{H}{$\beta$} emission at $z \sim 8.7$. We additionally note that we recover two of the NIRCam-selected, then NIRSpec-confirmed, $z = 8.7 - 9$ objects reported in the literature \citep[CEERS MSA IDs 2 and 7, our EGS-109389 and EGS-110201, respectively;][]{fujimoto2023, nakajima2023, arrabalharo2023b, harikane2023_specz}. The remaining three objects with spectroscopic redshifts reported in the literature (CEERS MSA IDs 23, 24, and 80083; \citealt{tang2023, fujimoto2023, arrabalharo2023b, harikane2023_specz}) are relatively faint, where the completeness of our selection is expected to be low.

\begin{figure*}
    \centering
    \includegraphics[width=\textwidth]{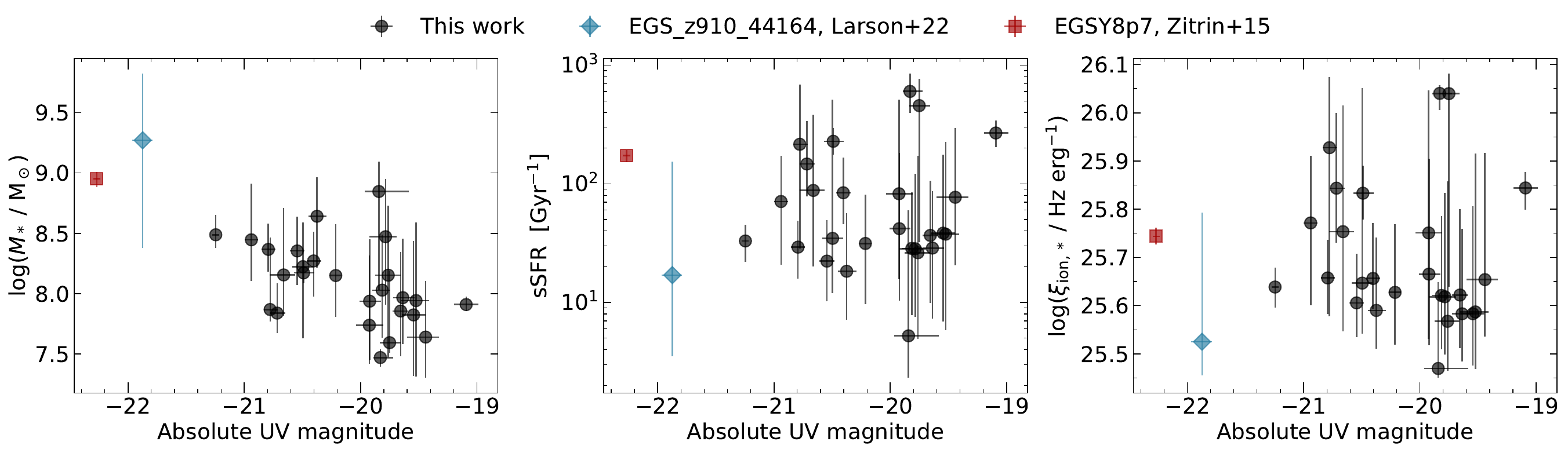}
    \caption{The inferred stellar masses, sSFRs, and intrinsic stellar ionizing photon production efficiencies (before stellar photons are processed through dust and gas) of our sample as a function of absolute UV magnitude with error bars showing formal uncertainties from the SED models, which may be a factor of a few larger when considering systematics introduced by e.g. altering the stellar IMF or model SFH. We also infer the properties of the two bright LAEs in the field, shown as colored points. The red square is EGSY8p7 \citep{zitrin2015} and the blue-green diamond denotes \larsonlae\ \citep{larson2022}. We note that we adopt the \textit{HST} and \textit{Spitzer} photometry reported by \citet{finkelstein2022} for \larsonlae, as it is not covered by the CEERS NIRCam imaging. The objects in our sample are much fainter than either of the LAEs and correspondingly, have smaller stellar masses. Our candidates also have relatively large sSFRs ($\sim 5 - 600$\,Gyr$^{-1}$ with median $42$\,Gyr$^{-1}$) and are very efficient ionizing agents ($\xi_{\text{ion}}^{*} = 10^{25.5 - 26.0}$\,Hz\,erg$^{-1}$ with median $\xi_{\text{ion}}^{*} = 10^{25.6}$\,Hz\,erg$^{-1}$).}
    \label{fig:properties}
\end{figure*}

\subsection{Photometric redshifts} \label{subsec:photozs}

We next wish to assess whether the candidates we have identified may lie at redshifts similar to that of the two known bright LAEs. We therefore infer the photometric redshifts of the objects in our sample with the BayEsian Analysis of GaLaxy sEds tool \citep[\beagle;][]{chevallard2016}. \beagle\ computes the stellar and nebular emission of galaxies by combining the latest version of the \citet{bruzual2003} stellar population synthesis models with the \textsc{cloudy} photoionization code \citep{ferland2013}; see description by \citet{gutkin2016}. The updated \citet{bruzual2003} models are underpinned by single star stellar isochrones computed by the PAdova and TRieste Stellar Evolution Code \citep[\textsc{parsec};][]{bressan2012, chen2015}. For these models, we adopt a \citet{chabrier2003} initial mass function (IMF) with mass range $0.1 - 300$\,M$_\odot$, an SMC dust prescription \citep{pei1992}, and the IGM attenuation model of \citet{inoue2014}. We place a uniform prior on redshift ranging from $z_\text{phot} = 0 - 15$ and assume a constant star formation history (CSFH) with a log-uniform prior on age ranging from 1\,Myr to the age of the Universe at the redshift under consideration. We adopt uniform priors on logarithmic stellar mass ($5 \leq \log(M_* / \text{M}_\odot) \leq 12$), $V$-band optical depth ($-3 \leq \log(\tau_\textsc{v}) \leq 0.7$), stellar metallicity ($-2.2 \leq \log(Z_* / \text{Z}_\odot) \leq 0.24$), and ionization parameter ($-4 \leq \log(U) \leq -1$). Finally, we assume that the total interstellar gas- and dust-phase metallicity is equal to the stellar metallicity with dust-to-metal mass ratio to $\xi_d = 0.3$, noting that \beagle\ self-consistently models the effects of depletion.

We find compelling $z \sim 8 - 10$ solutions for all objects but two in our colour-selected sample. EGS-25508 has no acceptable solutions at $z_\text{phot} \geq 6$, so we remove this object from our sample. EGS-64770 is best fit at $z_\text{phot} = 7.7$, and while we retain this object in our sample, we caution that it may be at a slightly lower redshift than we target in this work. For the remainder of the sample, the median $z_\text{phot} = 8 - 10$ SEDs inferred by our models generally fit the data well within uncertainties. If the SED model fits the observed photometry exactly within uncertainties, we would expect $\chi^2 = 9$ or $\chi^2 = 10$ depending on the filters available at the location of the object,\footnote{$\chi^2 = \sum (f_\text{obs} - f_\text{model})^2 / \sigma_\text{obs}^2$,  where $f_\text{obs}$ and $\sigma_\text{obs}$ are the observed flux and $1\sigma$ flux error, $f_\text{model}$ is the SED model flux, and the sum runs over all of the nine or ten available photometric bands.} and we find that the $z_\text{phot} = 8 - 10$ SEDs have $\chi^2$ values of $1.3 - 19.4$, indicating that we find reasonable fits at $z = 8 - 10$.

Our goal for these models is to assess the feasibility of our colour selected candidates lying at redshifts near to those of the known LAEs in the EGS field. We find that the large majority of our sample ($18 / 23$, or 80\,per\,cent of the objects without spectroscopic redshifts) have very large probabilities ($\geq 95$\,per\,cent) of lying at $z > 6$, and twelve of those also have large, $\geq 90$\,per\,cent probabilities of lying in the more narrow redshift range of $z = 8.2 - 9.2$. However, we acknowledge that in some cases, there are non-negligible probabilities of lower redshift ($z \sim 2$) solutions that are at times degenerate with the $z \sim 8 - 10$ solutions. These lower redshift solutions are often extremely dusty with extreme emission lines, with typical stellar masses of $M_* \approx 10^{7 - 8}$\,\Msun, typical ages of a few hundred Myr, stellar metallicities of $Z_* \sim 0.1 - 0.2$\,Z$_\odot$, and strong dust attenuation ($A_\textsc{v} \sim 0.7$, where $A_\textsc{v} = 1.086\tau_\textsc{v}$). These degeneracies underline the need for spectroscopic confirmation of photometric candidates to rigorously assess the neighbourhoods of bright LAEs during reionization. Nevertheless, we emphasize that, though these lower redshift solutions exist, we still find credible $z \sim 8 - 10$ SEDs for all candidates in the sample. Thus, we proceed with our entire colour selected sample to evaluate any potential overdensity associated with the LAEs.

\subsection{Sample properties} \label{subsec:sample}

The ionizing, star-forming, and morphological properties of the $z \sim 8.7$ galaxies in our sample grant insight into not only their potential contribution to reionization, but also how these early galaxies first assembled. We therefore wish to infer these physical properties from the photometric SEDs we have measured (see Figure\ \ref{fig:seds} for example SEDs from our sample). To do this, we use \beagle\ to re-fit the photometric data with a very similar model set up as our models designed to infer photometric redshifts for selection described in Section\ \ref{subsec:photozs}, but restrict the redshift to $z_\text{phot} = 4 - 12$ (uniform prior) and logarithmic stellar metallicity to $-2.2 \leq \log(Z / Z_\odot) \leq -0.3$ (uniform prior). We have set an upper limit on stellar metallicity of $Z \sim  0.5$\,Z$_\odot$ to avoid solutions with unreasonably high, near-solar metallicities for our $z \sim 8.7$ galaxies. For objects with known redshifts (EGSY8p7/EGS-69787, EGS-83141, EGS-87873, EGS-109389, and EGS-110201), we fix the redshift to the systemic redshift. We report selected physical properties inferred from these SED models, as well as selected observational properties, for each object in our sample in Table\ \ref{tab:sample_properties}.

The objects in our sample, excluding EGSY8p7, have observed F200W magnitudes ranging from $\text{F200W} = 26.2 - 28.0$ (median $\text{F200W} = 27.3$), implying that these objects are inherently moderately faint if they lie at our redshifts of interest, $z \sim 8.7$. We also find that most objects have moderately blue observed rest-UV continuum slopes\footnote{Calculated by fitting $f_\lambda \propto \lambda^{\beta}$ to the observed F150W, F200W, and F277W fluxes, i.e. the filters expected to probe the rest-UV that are not attenuated by the IGM at our redshifts of interest.} of $-2.9 \leq \beta \leq -1.3$ (median $\beta = -2.2$), consistent with expectations for $z \gtrsim 6$ galaxies \citep[e.g.][]{finkelstein2012, dunlop2013, bouwens2014, bhatawdekar2021, topping2022, cullen2023}. From the \beagle\ models described above, we infer absolute UV magnitudes ranging from $-21.2 \lesssim M_\textsc{uv} \lesssim -19.1$,\footnote{Calculated by integrating the \beagle\ model spectra over a value unity tophat from rest-frame wavelengths of $\lambda_\text{rest} = 1450 - 1550$\,\AA.} corresponding to $\sim (0.1 - 1) \times L_\textsc{uv}^*$ assuming a characteristic luminosity of $M_\textsc{uv}^* = -21.2$ \citep[approximately consistent with Schechter parametrizations of the $z \sim 9$ UV luminosity function derived from both \textit{HST} and \textit{JWST} data; e.g.][]{bouwens2021, harikane2023}. We infer $V$-band dust attenuations of $A_\textsc{v} = 0.01 - 0.4$ with median $A_\textsc{v} = 0.01$ from the \beagle\ models, suggesting most of these objects are minimally impacted by dust, as expected from the blue observed UV slopes.

At $z \sim 8.7$, the $\gtrsim 3$\,$\mu$m imaging from \textit{JWST} probes rest-frame optical emission (see SEDs in Figure\ \ref{fig:seds}), allowing us to gain detailed insights into the star-forming and ionizing properties of the sample. For example, if F356W, F410M, and F444W are all brighter than the bluer filters, this may empirically suggest the presence of a Balmer break built up by a mature stellar population. Indeed, one object in our sample (EGS-91290) is $\sim 0.4 - 0.8$\,mag brighter in F356W, F410M, and F444W than in the bands that probe the rest-UV continuum, and is best fit with an old CSFH age of $\sim 200$\,Myr (implying a formation redshift of $z \sim 12$ assuming the median photometric redshift of this object, $z_\text{phot} = 8.71$) and a moderately large stellar mass of $\sim 10^{8.8}$\,M$_\odot$ (assuming a CSFH).

Moreover, at $z \lesssim 8.9$, the strong rest-optical emission lines \ion{H}{$\beta$} and \ion{[O}{iii]}$\lambda\lambda$4959,5007 transmit through the F444W filter. Thus, a red $\text{F410M} - \text{F444W}$ color implies both the presence of strong \ion{[O}{iii]}+\ion{H}{$\beta$} emission and a redshift of $z \lesssim 8.9$. Specifically, at $z = 8.7$, $\text{F410M} - \text{F444W} = 0.5$ suggests $\text{EW}_{\text{[O}\,\textsc{iii]}+\text{H}\,\beta} \approx 640$\,\AA, slightly less than the median EW of $\sim 770$\,\AA\ expected for $z \sim 6.5 - 8$ galaxies at similar luminosities found by \citet{endsley2023_ceers}. In our sample, we find that nine of the 27 objects have $\text{F410M} - \text{F444W} \geq 0.5$, consistent with the very large \ion{[O}{iii]}+\ion{H}{$\beta$} EWs expected for this population \citep[e.g.][]{deBarros2019, endsley2021_OIII, endsley2023_ceers}. Of the remaining 18 objects, seven have $\text{F410M} - \text{F444W} = 0.3 - 0.5$, implying moderate \ion{[O}{iii]}+\ion{H}{$\beta$} emission $\text{EW}_{\text{[O}\,\textsc{iii]}+\text{H}\,\beta} \approx 350 - 640$\,\AA.

We next investigate whether this distribution of $\text{F410M} - \text{F444W}$ colors is consistent with expectations for \ion{[O}{iii]}+\ion{H}{$\beta$} emission in this population. Following the methods of \citet{gehrels1986} assuming binomial statistics, we estimate the fraction of the sample that we would expect to observe with $\text{F410M} - \text{F444W} < 0.5$ given the \ion{[O}{iii]}+\ion{H}{$\beta$} EW distribution for $M_\textsc{uv} \lesssim -19$ galaxies at $z \sim 6.5 - 8$ found by \citet{endsley2023_ceers}. To a $1\sigma$ confidence level, we find that we would expect to observe up to 14 of 27 objects with $\text{F410M} - \text{F444W} \lesssim 0.5$ if all are at $z \lesssim 8.9$, while we observe 18 in our sample. However, we note that we expect to observe more objects with $\text{F410M} - \text{F444W} \lesssim 0.5$ colours if any objects in our sample are at redshifts of $z \gtrsim 8.9$, as the stronger \ion{[O}{iii]}$\lambda$5007 component of the \ion{[O}{iii]}$\lambda\lambda$4959,5007 doublet quickly redshifts out of F444W at $z \gtrsim 8.9$. Thus, the large number of objects lacking strong F444W excesses could suggest that some of the objects in our sample may lie at slightly higher redshifts of $z \sim 8.9 - 9.1$, as would potentially be expected from our selection completeness (Figure\ \ref{fig:completeness}). Alternatively, it may be possible that more objects than expected are members of the population of weak \ion{[O}{iii]}+\ion{H}{$\beta$} emitters observed in photometric samples of galaxies with similar luminosities at $z \sim 6 - 8$ \citep[e.g.][]{endsley2023_ceers, endsley2023_jades}, as well as at least one spectroscopically confirmed at $z = 8.807$ with an \ion{[O}{iii]}$\lambda$5007 EW of $\sim 370$\,\AA\ \citep{fujimoto2023}. Such weak line emission could potentially suggest several physical situations: (1) a high occurrence rate of low metallicities ($Z \lesssim 0.01 - 0.05$\,Z$_\odot$, which have been observed in large numbers in photometric $z \gtrsim 6$ galaxy samples from \textit{JWST}; e.g. \citealt{endsley2023_ceers, endsley2023_jades}), leading to weaker \ion{[O}{iii]} emission than is found in more chemically evolved systems; (2) a large fraction of galaxies being observed after a rapid downturn in star formation rate (SFR) over the most recent $\sim 10$\,Myr, leading to a relatively small population of OB stars to produce strong nebular line emission; or (3) large Lyman continuum escape fractions, leading to a decrease in the interactions between hydrogen-ionizing photons and the interstellar medium that produce nebular lines \citep[e.g.][]{zackrisson2013}. For these weak line emitters, our \beagle\ models with a CSFH and fixed $f_\text{esc} = 0$ generally tend to find solutions at $z \sim 8.5 - 8.9$ (driven primarily by the strength of the observed \Lya\ break) with low metallicities rather than solutions at higher redshifts. More model flexibility in quantities such as the escape fraction or the SFH may be justified to better fit the large population of weak line emitters and could ameliorate the models' need to fit these systems with low metallicity solutions, but ultimately, upcoming future spectroscopy will be necessary to confirm the redshifts and emission line properties of these objects to precisely characterize their properties and spatial distribution.

\begin{figure}
    \centering
    \includegraphics[width=\columnwidth]{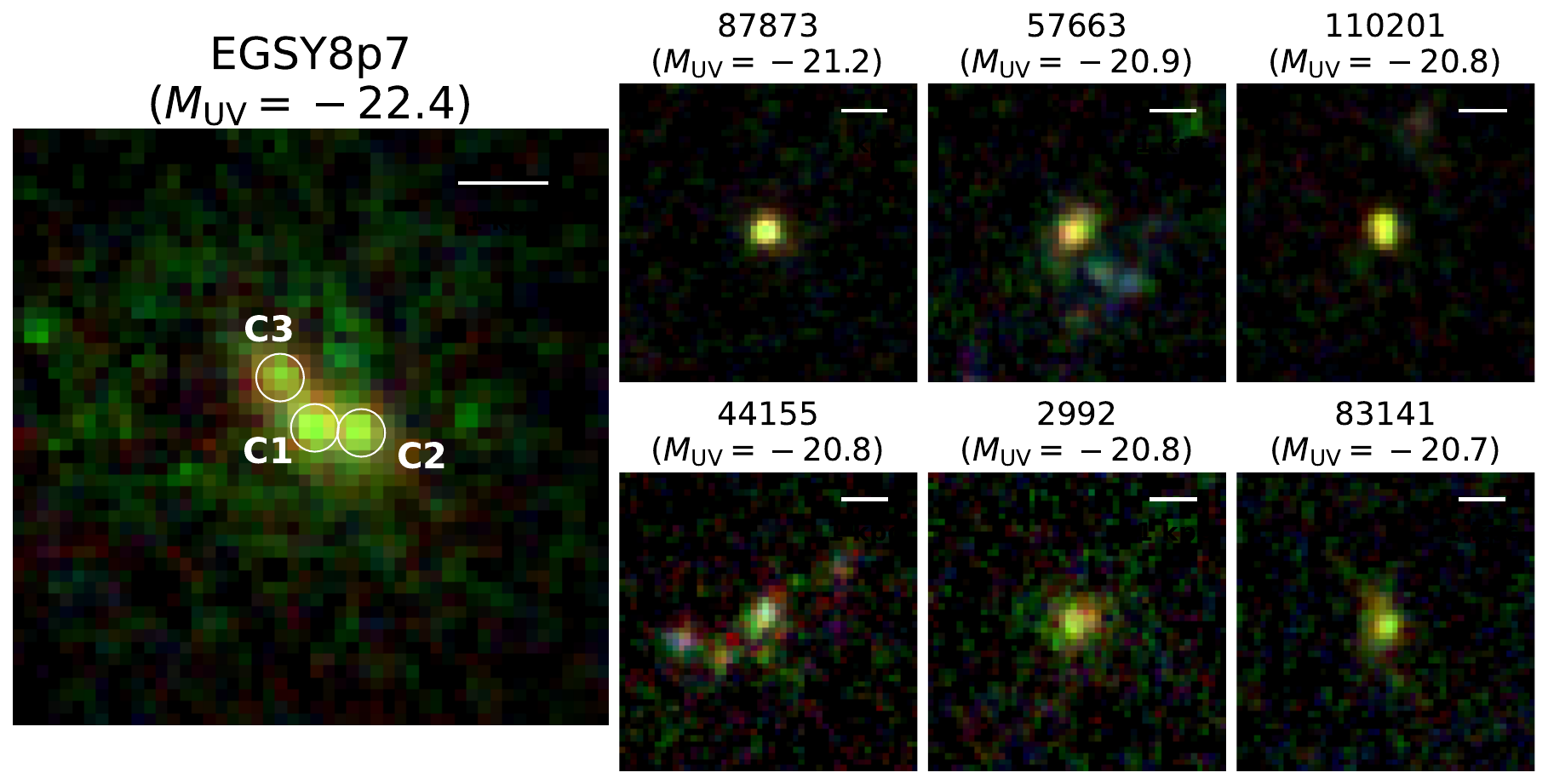}
    \caption{Rest-frame UV colour images (red = F200W, green = F150W, blue = F115W) of EGSY8p7 (left, large panel) and the six most luminous objects in our sample (remaining six, smaller panels) including three spectroscopically confirmed: EGS-87873 at $z = 8.715$ \citep{tang2023}, EGS-110201 at $z = 8.876$ \citep{fujimoto2023}, and EGS-83141 at $z = 8.763$ \citep{arrabalharo2023a}. Our sample is generally much fainter than EGSY8p7, with the brightest object being $\sim 1.2$\,mag fainter. We highlight the complex morphology of EGSY8p7, which has three distinct star-forming clumps, labelled C1, C2, and C3 in the image. In contrast, the majority of the fainter candidates in our sample have much simpler single-component morphologies.}
    \label{fig:morphology}
\end{figure}

We now turn to investigating insights provided by the NIRCam photometry into the physical properties of the objects in our sample. In general, we find properties consistent with expectations for $z \sim 9$ galaxies previously studied with \textit{HST} and \textit{JWST} (see Figure\ \ref{fig:properties} for a summary of the properties we infer). Under the assumptions of a CSFH, we find moderate stellar masses of $M_* \approx 10^{7.5 - 8.8}$\,\Msun\ (median $M_* \approx 10^{8.0}$\,M$_\odot$). As expected for the fainter luminosities we can probe with \textit{JWST}, these masses are smaller than the masses inferred for the two LAEs and other \textit{HST}-selected objects at the same redshifts \citep[e.g.][]{tacchella2022_hst}, but are broadly consistent with inferences for $z \gtrsim 6$ galaxies at similar luminosities studied with \textit{JWST} \citep[e.g.][]{endsley2023_ceers, bradley2022, leethochawalit2023, santini2023, furtak2023, whitler2023_ceers, robertson2023}. We also find that the objects in our sample generally appear to be dominated by recent bursts of star formation, with large CSFH-based specific star formation rates (sSFR; $\text{sSFR} \approx 5 - 600$\,Gyr$^{-1}$ with median 42\,Gyr$^{-1}$), and are very efficient ionizing agents ($\xi_{\text{ion}}^{*} \approx 10^{25.5 - 26.0}$\,Hz\,erg$^{-1}$, median $10^{25.6}$\,Hz\,erg$^{-1}$, where $\xi_{\text{ion}}^{*}$ is the intrinsic stellar ionizing photon production efficiency before processing through dust and gas). These values are similar to the large SED-inferred sSFRs and ionizing photon production efficiencies for galaxies at $z \sim 6 - 8$ \citep[e.g.][]{endsley2021_OIII, stefanon2022_colors, endsley2023_ceers}.

We note that throughout our SED modelling, we have assumed a \citet{chabrier2003} stellar IMF and a constant SFH. Adopting an alternative IMF could change the inferred stellar masses by $\sim 0.2 - 0.4$\,dex \citep{wang2023, woodrum2023} and increase uncertainties by a factor of $\sim 2-3$ over formal errors from SED models. An alternative form of the SFH model (e.g. a non-parametric model that allows significant stellar mass formation at early times) may lead to larger stellar masses than the CSFH model, typically by factors of up to $\sim \text{five}$ for objects such as those in our sample with CSFH ages of a few tens of Myr \citep[e.g.][]{whitler2023_cosmos, endsley2023_jades}. The form of the SFH may also alter the ionizing photon production efficiencies inferred; for example, a two-component parametric SFH \citep[a delayed-$\tau$ SFH at early times and a constant component at recent times that is decoupled from the delayed-$\tau$ model; see][]{endsley2023_ceers, endsley2023_jades} infers smaller ionizing photon production efficiencies by less than a factor of three for the majority our sample. Ultimately, these effects may combine to increase uncertainties by factors of a few over formal SED model uncertainties, though many of these model choices may have competing effects on the inferred galaxy properties and thus quantifying systematic effects will require a better understanding of e.g. the true IMF and the typical galaxy SFH during reionization.

In addition to the SED-based properties we infer, the high angular resolution of the \textit{JWST}/NIRCam imaging allows us to examine the morphological properties of the sample. As shown in Figure\ \ref{fig:morphology}, the majority of the faint objects we have identified in this work are less morphologically complex than the very bright EGSY8p7. They often have only one star-forming component while EGSY8p7 has three, consistent with findings from previous studies that clumpy morphologies become more common at brighter luminosities \cite[e.g.][]{bowler2017, chen2023}. Following the methods of \citet{chen2023}, we derive the half-light radii, $r_e$, of these objects by fitting their F150W surface brightness profiles at original resolution (i.e. before PSF homogenization) with fixed $n = 1$ S{\'e}rsic profiles (i .e. exponential profiles) using the \textsc{lmfit} package \citep{erwin2015}. For the three components of EGSY8p7, we assume an exponential profile for each component and fit them simultaneously, and find that each of its three star forming complexes are extremely compact ($r_e \lesssim 140$\,pc in the rest-UV, upper limit corresponding to the F150W pixel scale of 30\,mas) and are each separated by $\sim 900$\,pc. We also find that the single components of the faint galaxies in our sample are often relatively compact. Specifically, we measure half-light radii in the rest-UV of the five brightest single-component candidates in our sample range from $r_e \lesssim 140$\,pc to $\sim 650$\,pc. This implies star formation rate surface densities ($\Sigma_\textsc{sfr} = (\text{SFR} / 2) / (\pi r_e^2)$) ranging from $\Sigma_\textsc{sfr} \sim 6$\,\Msun\,yr$^{-1}$\,kpc$^{-2}$ up to $\Sigma_\textsc{sfr} \gtrsim 80$\,\Msun\,yr$^{-1}$\,kpc$^{-2}$, which are larger than observed in $z \sim 0$ galaxies \citep[e.g.][]{shibuya2015} but similar to those found for other systems at $z \sim 9$ \citep{ono2022}.

Using the resolved NIRCam imaging, we can also investigate the possibility for a spatially extended or offset component of old stars in EGSY8p7 itself, whose light would be outshined by any young stars present in the integrated photometry \citep[for a more detailed discussion of this effect in the general UV-bright population, see][]{chen2023}. If such a component were to exist, this would imply that EGSY8p7 may have been contributing to the beginning stages of the growth of an early ionized bubble. However, we find no clear evidence that any component of EGSY8p7 is dominated by a mature ($\gtrsim 100$\,Myr) stellar population. All three of the star-forming complexes labelled in Figure\ \ref{fig:morphology} have high ionizing production efficiencies of $\xi_{\text{ion}}^{*} = 25.8, 26.0,$ and $25.8$\,Hz\,erg$^{-1}$ for C1, C2, and C3, respectively, that suggest that EGSY8p7 could potentially ionize a large volume given long enough times. However, all three clumps also have extremely young CSFH ages of 4\,Myr, 2\,Myr, and 4\,Myr and correspondingly small stellar masses of $\log(M_* / \text{M}_\odot) \approx 7.7, 7.4,$ and $7.7$. This suggests that EGSY8p7 is dominated by stars formed in a recent burst of star formation, and if there is an older population of stars, they are outshined by the recently formed population.

However, though the rest-UV and optical SED is likely dominated by a young stellar population formed in a burst of star formation, it may be possible that the SED at longer wavelengths is more sensitive to the presence of a more mature stellar population. It has been demonstrated by \citet{papovich2022} that, relative to pre-\textit{JWST} constraints from \textit{Spitzer}/IRAC data, the addition of MIRI data covering the rest-optical SED of EGSY8p7 leads to significantly more precise constraints on the stellar mass. Here, we investigate whether the addition of longer wavelength MIRI data can significantly influence quantities such as the SFH and stellar mass when starting from the NIRCam SED for EGSY8p7. We note that the F410M NIRCam filter already provides a constraint on the rest-frame optical continuum ($\lambda_\text{rest} \approx 3980 - 4430$\,\AA), though MIRI probes much longer rest-frame wavelengths (up to $\lambda_\text{rest} \sim 8900$\,\AA\ with F770W).

To conduct this investigation, we repeat our \beagle\ SED modelling analysis assuming a CSFH, adding the MIRI F560W and F770W photometry reported by \citet{papovich2022} to the NIRCam SED. In general, we find similar properties with the MIRI data as without: CSFH-based stellar mass of $\log(M_* / \text{M}_\odot) \sim (8.8 - 9.0) \pm 0.1$ and age of $\sim 4 - 6$\,Myr, suggesting that the NIRCam imaging probing the rest-optical continuum may be sufficient to constrain the mass of the stellar population dominating the SED. However, we also investigate whether the addition of MIRI photometry enables us to place stronger upper limits on the early SFH than is possible with NIRCam alone. If so, it may be possible to more precisely quantify the contribution of ionizing photons from early star formation in EGSY8p7 towards the early growth of an ionized bubble.

\begin{figure}
    \centering
    \includegraphics[width=\columnwidth]{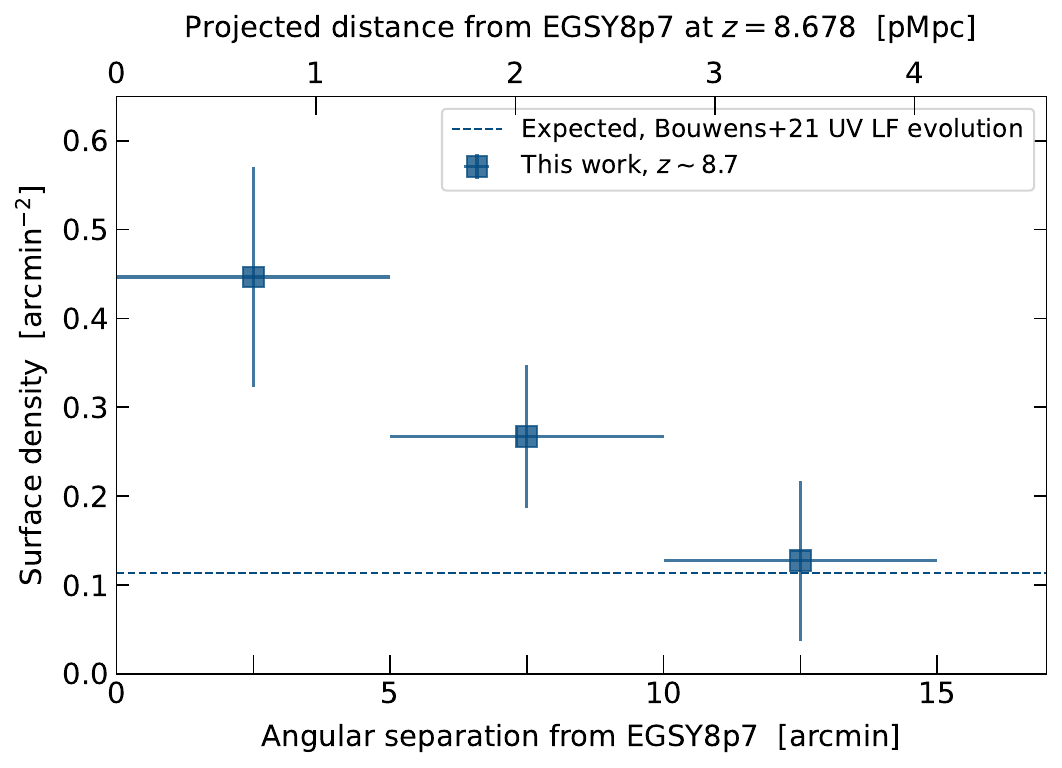}
    \caption{The observed surface density of the objects in our sample as a function of angular separation from EGSY8p7. On the top $x$-axis, we convert this angular separation to a projected physical separation from EGSY8p7 at its systemic redshift, $\zsys = 8.678$ \citep{tang2023, larson2023}. At the closest separations we consider, $< 5$\,arcmin (corresponding to $\sim 1.4$\,pMpc at $z = 8.678$), we find a surface density of $(0.45 \pm 0.12)$\,arcmin$^{-2}$, a factor of $\sim \text{four}$ overdense relative to expectations derived from the evolution of the UV LF \citep{bouwens2021} after accounting for the completeness of our selection. This overdensity decreases with increasing distance away from EGSY8p7 to a factor of $\sim 1.9$ overdense at separations of $5 - 10$\,arcmin ($\sim 1.4 - 2.7$\,pMpc) until the densities are consistent within Poisson uncertainties with an average field at separations of $10 - 15$\,arcmin ($\sim 2.7 - 4.1$\,pMpc).}
    \label{fig:overdensity}
\end{figure}

To this end, we model the NIRCam and MIRI SED of EGSY8p7 with a non-parametric SFH model using a prior that not only allows, but tends to prefer an extended early period of star formation. We follow the methods of \citet{whitler2023_cosmos}, assuming the non-parametric `continuity' prior of the Bayesian SED modelling tool \prospector\ \citep{leja2019, johnson2021} and adopting eight age bins for the SFH, with the two most recent age bins fixed to $0 - 3$\,Myr and $3 - 10$\,Myr and the rest spaced evenly in logarithmic time. With this model, we infer a factor-of-six larger stellar mass ($\log(M_* / \text{M}_\odot) \sim 9.7$) than the CSFH model, both with and without MIRI. Ultimately, this suggests that for objects similar to EGSY8p7, with very young light-weighted ages that suggest a recent burst of star formation, the SED continues to be dominated by nebular continuum emission well into the rest-frame optical. Thus, even with photometry probing wavelengths as red as $\lambda_\text{rest} \sim 8900$\,\AA, significant systematic uncertainties in the stellar masses of burst-dominated objects can still exist, which may ultimately require rest--near-IR observations or dynamical masses to resolve \citep[e.g.][]{tang2022_sfhs}.

This stellar mass difference is caused by a long, early period of early star formation ($\text{SFR} \sim 10$\,\Msun\,yr$^{-1}$, leading to a half-mass age of $\sim 160$\,Myr) in the non-parametric model, which suggests that EGSY8p7 could have feasibly been contributing to the reionization process for several hundred Myr. Moreover, there is evidence that EGSY8p7 may host an active galactic nucleus (AGN) \citep{mainali2018, larson2023}, which could have undergone past periods of rapid growth in the past despite being a sub-dominant component of the rest-UV and optical emission at the time of observation. In a similar manner as past star formation activity, past AGN activity could also contribute copious amounts of ionizing photons towards the growth of a large ionized bubble at early times \citep{larson2023}, though such activity may also be short-lived or episodic.

\begin{figure*}
    \centering
    \includegraphics[width=\textwidth]{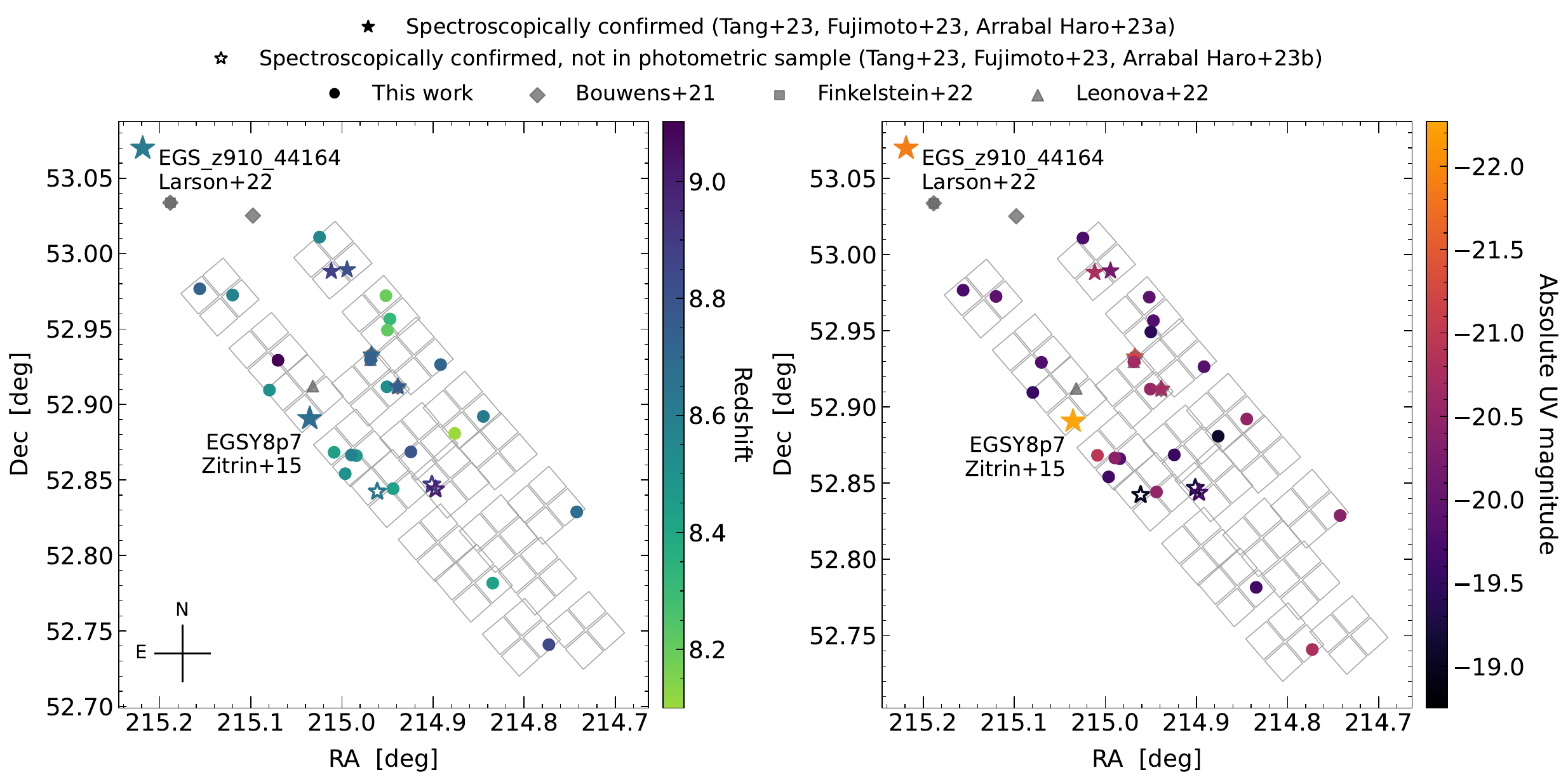}
    \caption{The on sky distribution of the candidates identified by our colour selection criteria (Section\ \ref{sec:selection}), along with photometric candidates previously identified from \textit{HST} searches over EGS with reported photometric redshifts of $z_\text{phot} = 8.2 - 9.2$ \citep[grey points;][]{bouwens2021, finkelstein2022, leonova2022}, and three spectroscopically confirmed galaxies at $z_\text{spec} = 8.881, 8.998,\text{ and } 8.638$ that are not included in our photometric sample \citep[open stars;][]{tang2023, fujimoto2023, arrabalharo2023b}. Objects from the literature are shown as grey diamonds \citep{bouwens2021}, squares \citep{finkelstein2022}, and triangles \citep{leonova2022}. We note that we do not show objects from the literature if the NIRCam photometry indicates they do not lie in our targeted redshift range; see the discussion in Section\ \ref{subsec:selection}. The CEERS NIRCam imaging footprint is outlined in grey. Finally, the candidates we identify in this work are shown as filled, coloured points -- stars if they are spectroscopically confirmed \citep{tang2023, fujimoto2023, arrabalharo2023a} and circles if they are not -- and the two bright $z = 8.7$ LAEs \citep{zitrin2015, larson2022} are shown as large stars. In the left panel, we colour our sample by the redshift, either the median photometric redshift inferred by our SED models (see Section\ \ref{subsec:sample}) or the systemic spectroscopic redshift if known. In the right panel, the colour bar shows the object's absolute UV magnitude inferred from our SED models. Besides EGSY8p7 itself, we identify 26 candidates over the entire field (ten of which are within the $\sim 20$\,arcmin$^2$ of the imaging that lies within 4\,arcmin of EGSY8p7), many more than the ten expected for an average field. Furthermore, most of the sample lies in the northeastern region of the footprint, relatively close to EGSY8p7 and \larsonlae.}
    \label{fig:map}
\end{figure*}

\section{A galaxy overdensity near UV luminous \texorpdfstring{\ion{L\lowercase{y}}{$\alpha$}}{Lya} emitters at \texorpdfstring{$\lowercase{z} = 8.7$}{z = 8.7}} \label{sec:overdensity}

With our sample of galaxies selected to lie near $z = 8.7$, we now turn to evaluating the potential for a pMpc-scale overdensity and ionized bubble associated with the two $z = 8.7$ LAEs in the EGS field. Using our NIRCam-based sample and SEDs, we can extend previous \textit{HST}-based investigations of the environments of the LAEs \citep[e.g.][]{finkelstein2022, leonova2022} to both fainter luminosities and larger physical scales. The depth of the NIRCam imaging allows us to identify objects as faint as $M_\textsc{uv} \sim -19$ ($\sim 0.1 L_\textsc{uv}^*$), probing the faint galaxy population thought to provide the bulk of ionizing photons needed to create an ionized bubble. Furthermore, the imaging extends up to $\sim 15$\,arcmin away from EGSY8p7 and $\sim 27$\,arcmin away from \larsonlae\ (physical separations up to $\sim 4$\,pMpc and $\sim 7$\,pMpc in projection at $z = 8.7$, respectively), allowing us to search for faint galaxies over pMpc scales that may trace correspondingly large ionized bubbles. We concentrate primarily on the galaxy population near EGSY8p7, as \larsonlae\ does not fall in the CEERS NIRCam imaging footprint, but also comment briefly on the distribution of galaxies between the two LAEs. We focus on the sample colour selected as described in Section\ \ref{subsec:selection}, as the inclusion of the objects identified by our LAE-targeted selection (Appendix\ \ref{appendix:lae_selection}) does not significantly impact our conclusions.

To quantify the amplitude and spatial extent of any overdensity that may exist, we first must determine the number of objects expected in an average field given our selection function and the UV luminosity function (UV LF). To achieve this, we assume a Schechter parametrization of the UV LF and the \citet{bouwens2021} redshift evolution of the Schechter parameters, then convolve with our analytic selection completeness (described in Section\ \ref{subsec:selection} and shown in Figure\ \ref{fig:completeness}). We find an average surface density of $0.11$\,arcmin$^{-2}$ in the redshift and magnitude range probed by our selection.\footnote{Assuming the UV LF parametrizations and $z \sim 9$ parameters derived by other studies \citep[e.g.][]{harikane2023, donnan2023} implies generally similar surface densities ranging from $\sim 0.07 - 0.12$\,arcmin$^{-2}$.} Thus, we would expect to identify ten objects over the entire $\sim 85$\,arcmin$^{2}$ field. However, we have identified ten candidates within only a \textit{four arcminute radius} of EGSY8p7 ($\sim 20$\,arcmin$^2$ of the imaging area), or $R \sim 1$\,pMpc in projection at $z = 8.7$, potentially suggesting at least a mild overdensity in the close vicinity of EGSY8p7.

We can now compare this expected surface density to the observed surface density implied by our sample in the field of the $z = 8.7$ Ly$\alpha$ emitters. As we are interested in assessing overdensities that may be associated with EGSY8p7, we calculate the surface densities in concentric circular annuli centred on EGSY8p7 with radii increasing in steps of $\Delta R = 5$\,arcmin (corresponding to $\Delta R = 1.4$\,projected\,pMpc at $z = 8.7$). These annuli intersect with 30\,arcmin$^2$, 41\,arcmin$^2$, and $16$\,arcmin$^2$ of the CEERS imaging area for separations $< 5$\,arcmin, $5 - 10$\,arcmin, and $10 - 15$\,arcmin, respectively. In marked contrast to the 0.11\,arcmin$^{-2}$ surface density expected for an average field, we find a surface density of $(0.45 \pm 0.12)$\,arcmin$^{-2}$ (uncertainties calculated assuming Poisson statistics) in the 30\,arcmin$^2$ of the imaging that lies within $R = 5$\,arcmin of EGSY8p7 -- a factor of four overdensity in projection. As the distance from EGSY8p7 increases, the surface density decreases to $(0.30 \pm 0.08)$\,arcmin$^{-2}$ at separations of $R = 5 - 10$\,arcmin, then $(0.13 \pm 0.09)$\,arcmin$^{-2}$ at $R = 10 - 15$\,arcmin (consistent with the surface density of an average field).\footnote{Including the objects identified by the LAE-targeted selection increases these observed surface densities by $\sim 0.1 - 0.2$\,arcmin$^{-2}$ at all separations.} We show these observed surface densities in comparison with the expected surface density in Figure\ \ref{fig:overdensity}. In particular, we highlight that, though the region within 5\,arcmin of EGSY8p7 comprises only a third of the total imaging area, nearly half of our photometric candidates (including two of the four spectroscopically confirmed objects) lie within that area; see Figure\ \ref{fig:map}. These objects are all significantly fainter and inferred to be much less massive than EGSY8p7. They also all have large sSFRs ($\sim 20 - 600$\,Gyr$^{-1}$) and large intrinsic ionizing photon production efficiencies ($\xi_\text{ion}^* \sim 10^{25.6 - 26.0}$\,Hz\,erg$^{-1}$) that are typical of the full sample (see Section\ \ref{subsec:sample}).

While the NIRCam imaging does not probe the volume immediately surrounding \larsonlae, we can use the imaging between EGSY8p7 and \larsonlae\ to qualitatively assess whether there is any evidence for a single large ionized region containing both LAEs. If the LAEs do inhabit the same ionized bubble, their $\sim 4$\,pMpc separation would require an ionized region with a radius of $R \sim 2$\,pMpc, at minimum. We note that, in a spherical geometry, this would place the LAEs very close to the edges of the bubble, near intergalactic \ion{H}{i} where \Lya\ may be easily attenuated. This would suggest that an even larger ionized region ($R \gtrsim 3$\,pMpc), at least along the line of sight, may be necessary to truly facilitate the transmission of \Lya\ through the IGM. However, even an $R = 2$\,pMpc ionized bubble would be larger than theoretically expected given the large neutral fraction expected at $z \sim 9$ \citep[e.g.][]{lin2016, geil2016, lu2023}, so we consider an $R = 2$\,pMpc ionized bubble as the most conservative case.

If there is one large ionized region encompassing both LAEs, we may expect a larger surface density of galaxies between the LAEs than in the areas furthest away from them. Indeed, we find that $\sim 80$\,per\,cent of the sample falls in the five most northeastern NIRCam pointings of the CEERS imaging, approximately between EGSY8p7 and \larsonlae\ ($\sim 6 - 15$\,arcmin away from \larsonlae, or $\sim 1.7 - 4$\,pMpc in projection at $z = 8.7$), while only three are in the most southwestern region of the imaging, $\sim 20 - 25$\,arcmin ($\sim 6 - 7$\,projected\,pMpc) away from \larsonlae\ (see Figure\ \ref{fig:map}). At face value, this could suggest a single large, $R \gtrsim 2$\,pMpc-scale ionized region encompassing both EGSY8p7 and \larsonlae. Such a structure could also potentially extend along the line of sight, as our photometric selection extends up to $z \sim 9$ and the fraction of objects with weak F444W excesses in our sample may imply that at least some of them could lie at $z \sim 8.9 - 9.1$ ($\sim 5 - 10$\,pMpc along the line of sight from $z = 8.7$); see discussion in Section\ \ref{subsec:sample}.

To investigate the plausibility of the presence of a large-scale ionized bubble, we estimate the size of the \ion{H}{ii} region that could be created by the galaxy population we have explored in this work. Adopting the methods described by \citet{endsley2022_bubble} \citep[see also][]{shapiro1987, haiman1997, cen2000}, we approximate the radius of a spherical \ion{H}{ii} region, $R_{\ion{H}{ii}}$, that a given population of galaxies within $R = 2$\,pMpc of EGSY8p7 (corresponding to a sphere of volume 33.5\,pMpc$^3$) could create using the following equation:

\begin{equation} \label{eqn:dR_dt}
    \frac{\text{d}R_{\ion{H}{ii}}}{\text{d}t} = \frac{\xi_\text{ion} f_\text{esc} L_\textsc{uv}}{4\pi R_{\ion{H}{ii}}^2 \langle n_{\ion{H}{i}}(z) \rangle} + R_{\ion{H}{ii}} H(z) - R_{\ion{H}{ii}} \alpha_\textsc{b} \langle n_{\ion{H}{i}}(z) \rangle \frac{C}{3}.
\end{equation}
The first two terms of Equation\ \eqref{eqn:dR_dt} describe the growth of ionized bubbles due to photoionizations from the ionizing sources and the Hubble flow, respectively, while the last term accounts for recombinations in the IGM. At the redshift, $z$, corresponding to the time step under consideration, $H(z)$ is the Hubble parameter and $\langle n_{\ion{H}{i}}(z) \rangle$ is the average neutral hydrogen density. $\alpha_\textsc{b}$ is the case B recombination coefficient, where we have assumed a temperature of $10^{4}$\,K, leading to a value of $\alpha_\textsc{b} = 2.59 \times 10^{-13}$\,cm$^{3}$\,s$^{-1}$ \citep{osterbrock2006}. $C = \langle n_{\ion{H}{i}}^2 \rangle / \langle n_{\ion{H}{i}} \rangle^2$ is the `clumping factor' that accounts for inhomogeneities in the IGM \citep{madau1999}. Motivated by theoretical expectations from simulations for the average IGM \citep[e.g.][]{shull2012, finlator2012, kaurov2015, gorce2018}, we assume a fixed value of $C = 3$, though we note that if the region is overdense, the clumping factor may be higher \cite[e.g.][]{mao2020, bianco2021}, leading to smaller bubbles due to the increased recombination rate; we briefly present predicted bubbles sizes with clumping factors of $C = 10$, $C = 30$, and $C = 50$ in Appendix\ \ref{appendix:clumping_factor}. For $\xi_\text{ion}$, we assume the median ionizing photon production efficiency of our sample inferred from our SED models, $\xi_\text{ion} = 10^{25.7}$\,Hz\,erg$^{-1}$. We note that this quantity is the ionizing photon production efficiency after the processing of stellar photons through dust and gas has occurred, in contrast to the previously reported intrinsic stellar ionizing photon production efficiency, $\xi_\text{ion}^*$. To calculate the non-ionizing UV luminosity, $L_\textsc{uv}$, we again assume the same Schechter parametrization of the UV LF as we adopted to estimate the expected surface density in the field, and consider the contributions of all galaxies $M_\textsc{uv} = -15$ and brighter \citep[approximately the faint end turnover of the UV LF tentatively measured in lensed fields,][]{bouwens2022_fflf}. To obtain $L_\textsc{uv}$ in a field that is a factor of $N$ overdense, we assume that the overall shape of the UV LF is the same as that in an average field, but with a higher normalization by a factor of $N$ (i.e. we multiply the normalization of the Schechter function by $N$). Finally, we integrate Equation\ \eqref{eqn:dR_dt} over time assuming that all galaxies in the region have been producing ionizing photons at the same rate with the same escape fraction for their entire lifetime.

\begin{figure}
    \centering
    \includegraphics[width=\columnwidth]{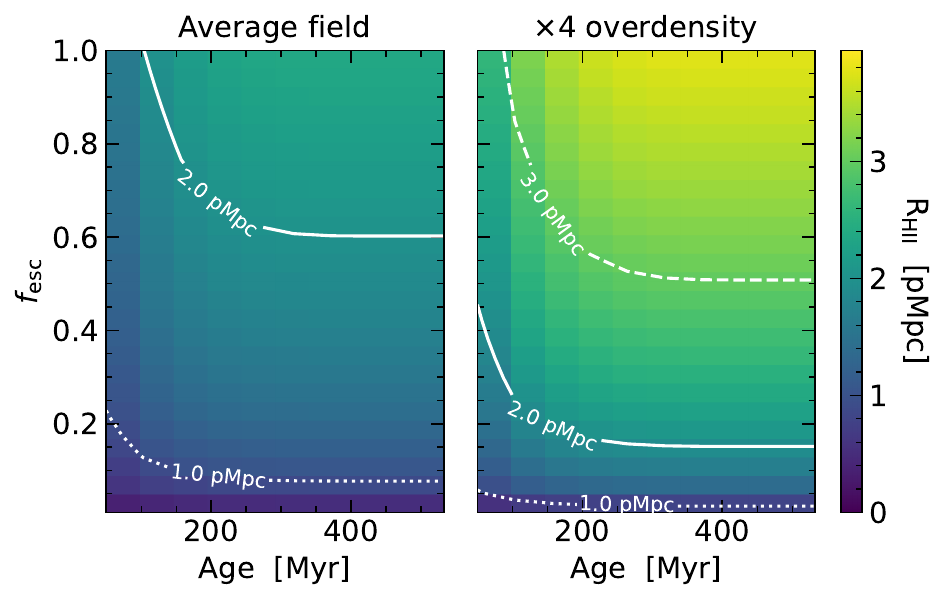}
    \caption{Predicted sizes of the ionized bubbles that could be created by the $M_\textsc{uv} \leq -15$ galaxy population within a 33.5\,pMpc$^3$ volume, corresponding to a sphere of radius $R = 2$\,pMpc that would encompass both $z = 8.7$ LAEs in the EGS field. The colourbar shows the predicted radius of a spherical ionized region, $R_{\ion{H}{ii}}$, as a function of the length of time the galaxies have been producing ionizing photons (i.e. the age) and the escape fraction of hydrogen-ionizing Lyman continuum photons ($f_\text{esc}$). We assume an ionizing photon production efficiency of $\xi_\text{ion} = 10^{25.7}$\,Hz\,erg$^{-1}$ and show the predicted bubble sizes if the number density is that of an average volume in the left panel, and if the volume is a factor of four overdense in the right panel. To ionize an $R \geq 2$\,pMpc spherical volume, galaxies in an average field would need very high escape fractions that are rarely seen even in individual objects with Lyman continuum detections. However, a galaxy population that is a factor of four overdense may be able to power an $R \geq 2$\,pMpc bubble with only moderate escape fractions ($f_\text{esc} \sim 0.15$) if they are producing ionizing photons for several hundred million years.}
    \label{fig:R_HII}
\end{figure}

Figure\ \ref{fig:R_HII} shows the predicted radii of \ion{H}{ii} regions for a variety of ages and escape fractions of ionizing photons ($f_\text{esc}$) assuming an average field (left panel) and a field that is a factor of four overdense (right panel). Under the assumptions described above, we conclude that for $M_\textsc{uv} \leq -15$ galaxies in an average field to power an ionized bubble large enough to encompass both LAEs within the age of the Universe, they would all require escape fractions of $f_\text{esc} \gtrsim 0.6$, significantly larger than expected for the population \citep[and are rare even in individual objects with Lyman continuum detections at $z \lesssim 3$; e.g.][]{izotov2018, pahl2021, flury2022}.

However, if the volume hosts a factor of four galaxy overdensity, $M_\textsc{uv} \leq -15$ galaxies may be able to create an $R \geq 2$\,pMpc bubble with an escape fraction of $f_\text{esc} \approx 0.15$ over $\gtrsim 200$\,Myr (approximately a third of the age of the Universe at $z = 8.7$), or with slightly higher escape fractions over slightly shorter times (e.g. $f_\text{esc} \approx 0.25$ for 100\,Myr). Such escape fractions, while somewhat high, are found in lower redshift samples and are consistent with inferences from indirect tracers for $f_\text{esc}$ for galaxies during reionization \citep[e.g.][]{meyer2020}. Ultimately, this suggests that the galaxy population we have considered in this work, combined with the fainter galaxies under our detection limit, may be able to create a large enough ionized region to contain both LAEs over long times.

However, we highlight that though the escape fractions may be physically plausible, the galaxy population we have considered in this work would require upwards of a few hundred Myr to carve out an $R \geq 2$\,pMpc bubble. From our CSFH models, we have inferred that the SEDs of our sample generally tend to be dominated by much shorter, recent bursts of star formation. Specifically, we have inferred a median CSFH sSFR of$\sim 42$\,Gyr$^{-1}$, which implies star formation on timescales of only a few tens of Myr (Section\ \ref{subsec:sample}) -- insufficient time to carve out an $R \geq 2$\,pMpc bubble under our assumptions for Equation\ \eqref{eqn:dR_dt}. If such a bubble does exist, this could imply that the amplitude of the galaxy overdensity around EGSY8p7 is stronger than a factor of four (which may be possible if most or all of the photometric candidates in this work are at redshifts very close to $z = 8.7$). It could also imply galaxies fainter than $M_\textsc{uv} = -15$ are contributing to the process of creating the bubble, or that there is simply a larger population of faint galaxies (i.e. the faint-end slope of the UV LF is steeper than we have assumed here). Alternatively, we have shown that a much older stellar population formed during an extended early period of star formation can be hidden under the young stellar population dominating the SED for burst-dominated objects such as those in this work \citep[Section\ \ref{subsec:sample}; see also][]{tacchella2022_hst, topping2022_ssfrs, whitler2023_cosmos}. Such an extended early SFH, if it lasted for several hundred Myr, could enable the creation of a large ionized bubble, and may in fact be necessary for the objects in our sample to carve out an $R \geq 2$\,pMpc bubble. We note that the average SFH from a variety of simulations \citep[e.g.][]{dawoodbhoy2018, ocvirk2020, legrand2022} are qualitatively consistent with an early period of increasing star formation activity. Thus, we may indeed expect galaxies to have started creating a large ionized bubble at relatively early times, though a detailed accounting of ionized bubble growth would require careful tracking of $f_\text{esc}$, $\xi_\text{ion}$, and $M_\textsc{uv}$ over time, so we defer a comprehensive examination of the impact of the SFH on ionized regions to future work.

If an $R \geq 2$\,pMpc ionized bubble exists, it may be a rare and individual occurrence. Theoretical predictions by \citet{lu2023} suggest that even an $R = 2$\,pMpc bubble is very unlikely, even if the region hosts a factor of $3 - 5$ photometric galaxy overdensity (consistent with both the findings of \citealt{leonova2022} and the overdensity we measure in the inner $\sim 1.4$\,projected\,pMpc around EGSY8p7). Alternatively, it could be indicative that the reionization topology at $z = 8.7$ is characterized by larger bubbles than expected. We highlight that the small probability of finding a large ionized region at $z = 8.7$ is, in large part, due to the very large volume-averaged fraction of intergalactic neutral hydrogen expected at $z \sim 9$ \citep[$\overline{x}_{\ion{H}{i}} \sim 0.8 - 0.9$;][]{mason2019_Ndotion, ocvirk2021, rosdahl2022, lu2023, mitra2023, mutch2024}. If the IGM neutral fraction is lower, then ionized bubbles are expected to be larger given the same dominant source population of ionizing photons. Alternatively, at fixed neutral fraction, ionized bubbles can be larger (though also more rare) if brighter, more massive galaxies are the dominant source of ionizing photons \citep[e.g.][]{mesinger2016, geil2016, lu2023}. If an $R \geq 2$\,pMpc ionized bubble exists at all, it would facilitate significant transmission of \Lya\ through the IGM for both the known bright LAEs and their faint neighbours \citep[of which there may be $> 2 - 4$ within a $2 \times 2$\,arcmin$^2$ area around the central LAE;][]{qin2022}. Thus, upcoming deep \textit{JWST}/NIRSpec observations (\textit{JWST} GO-4287; PI: Mason) aimed at characterizing the ionizing properties, spatial distribution, and \Lya\ line properties (e.g. velocity offsets) of any faint LAEs that may exist in this volume will be important to distinguish these scenarios.

\section{Summary} \label{sec:summary}

In this work, we have used \jwst/NIRCam imaging from the CEERS program \citep[][Finkelstein et al. in prep]{finkelstein2023_ceers} in the EGS field to investigate the potential for an overdensity of faint galaxies in the vicinity of two extremely UV-luminous LAEs at $z = 8.7$. We have built a sample of photometrically selected galaxy candidates, quantified the 2D galaxy overdensity suggested by these objects, and examined the implications of such an overdensity for the existence of large ionized bubbles at $z \sim 9$. We summarize our key results below.

\begin{enumerate}
    \item We have developed a S/N and colour selection designed to identify moderately faint galaxies at $z \sim 8.4 - 9.1$, redshifts similar to those of the two known bright LAEs in the EGS field. Using this selection, we have identified 27 objects, including the one LAE that falls within the CEERS imaging footprint, EGSY8p7.
    \item We infer the physical properties of the sample with SED models and find that, excluding EGSY8p7, these objects are relatively faint with absolute UV magnitudes of $-21.2 \lesssim M_\textsc{uv} \lesssim -19.1$, corresponding to $\sim (0.1 - 1) \times L_\textsc{uv}^*$, and stellar masses of $7.5 \lesssim \log(M_* / \text{M}_\odot) \lesssim 8.8$. They are also very efficient ionizing agents ($\xi_{\text{ion}}^{*} = 10^{25.5 - 26.0}$\,Hz\,erg$^{-1}$).
    \item The faint objects in our sample typically have simple morphologies. They are often composed of only one star-forming clump, in contrast to the very bright EGSY8p7 ($M_\textsc{uv} \approx -22.4$), which has three very compact star-forming components. The single components in our sample are also typically compact, with rest-UV half-light radii ranging from $r_e \lesssim 140$\,pc to 650\,pc. This implies star formation rate surface densities of at least $\Sigma_\textsc{sfr} \sim 6$\,M$_\odot$\,yr$^{-1}$\,kpc$^2$, and as high as $\Sigma_\textsc{sfr} \gtrsim 80$\,M$_\odot$\,yr$^{-1}$\,kpc$^2$ (adopting the SFRs inferred from our SED models).
    \item Besides EGSY8p7 itself, we have identified ten candidates within only $R = 4$\,arcmin of EGSY8p7 ($\sim 20$\,arcmin$^2$ of the CEERS imaging area) and 26 candidates over the full $\sim 85$\,arcmin$^2$ of the imaging. In contrast, for an average field of the same area, we would expect to identify only ten galaxies over the entire field. Compared to the expected surface density derived from the evolution of the UV luminosity function \citep{bouwens2021}, we find a factor of four overdensity within 5\,arcmin ($\sim 1.4$\,projected\,pMpc at $z = 8.7$) of EGSY8p7. At increasing separations from the LAE, the surface density of our sample decreases until it is consistent with an average field at $10 - 15$\,arcmin ($\sim 2.7 - 4.1$\,projected\,pMpc).
    \item Qualitatively, the spatial distribution of galaxies that we have identified over the field may suggest the existence of a large, $R \geq 2$\,pMpc bubble that contains both bright LAEs in the field. However, such an ionized bubble may be larger than theoretically expected. We investigate the properties that would be required for a population of galaxies to power an $R \geq 2$\,pMpc ionized bubble and find that if the $R = 2$\,pMpc spherical volume around EGSY8p7 is a factor of four overdense, the $M_\textsc{uv} \leq -15$ galaxy population could create this bubble over $\gtrsim 200$\,Myr with moderate escape fractions of $f_\text{esc} \approx 0.15$. Alternatively, the same galaxy population could power the same ionized bubble in less time, but would require slightly higher escape fractions (e.g. $f_\text{esc} \gtrsim 0.25$ over 100\,Myr).
    \item The SEDs of our sample tend to be dominated by light from stars formed in a recent burst of star formation over timescales of a few tens of Myr, shorter than the timescales required for the creation of a large, $R \geq 2$\,pMpc bubble. This could imply that extended, early SFHs may be necessary in order for the galaxy population we have considered in this work to power a large, pMpc-scale bubble.
    \item If a large, $R \geq 2$\,pMpc bubble at $z \sim 9$ exists, it may be a rare occurrence or could suggest that ionized regions at $z \sim 9$ may be larger than expected. This could potentially be explained if the IGM neutral fraction at $z \sim 9$ is lower than current models assume ($\overline{x}_{\ion{H}{i}} \sim 0.8 - 0.9$), or if the reionization process is dominated by the rare population of bright, massive galaxies that create larger ionized bubbles \citep[see also][]{lu2023}.
    \item Upcoming follow-up spectroscopy of the faint galaxies discussed in this work will be important to better constrain the properties of any ionized bubble, or bubbles, that may be associated with the LAEs. Spectroscopic observations will first enable us to confirm the existence of a galaxy overdensity associated with the LAEs. Then, detailed measurements of \Lya\ and other rest-UV and optical emission lines will then allow us to study the topology of the bubble(s) in detail and infer the properties of the ionizing galaxies to gain insights into the mechanisms by which the Universe is reionized.
\end{enumerate}

\section*{Acknowledgements}

The authors thank the anonymous referee for their constructive comments that improved this paper. We also thank Gabe Brammer for providing the optical imaging of the EGS as part of the CHArGE program and Jacopo Chevallard for access to the \beagle\ SED modelling tool used throughout this work. This work is based on observations made with the NASA/ESA \textit{JWST} and the data were obtained from the Mikulski Archive for Space Telescopes at the Space Telescope Science Institute, which is operated by the Association of Universities for Research in Astronomy, Inc., under NASA contract NAS 5-03127 for \textit{JWST}. These observations are associated with program number ERS-1345, and the authors thank the CEERS team led by Steven Finkelstein for their work developing their observing program.

LW acknowledges support from the National Science Foundation Graduate Research Fellowship under Grant No. DGE-2137419. DPS acknowledges support from the National Science Foundation through the grant AST-2109066. CM acknowledges support by the VILLUM FONDEN under grant 37459 and the Carlsberg Foundation under grant CF22-1322. The Cosmic Dawn Center (DAWN) is funded by the Danish National Research Foundation under grant DNRF140. This material is based on High Performance Computing (HPC) resources supported by the University of Arizona TRIF, UITS, and Research, Innovation, and Impact (RII) and maintained by the UArizona Research Technologies department.

We respectfully acknowledge the University of Arizona is on the land and territories of Indigenous peoples. Today, Arizona is home to 22 federally recognized tribes, with Tucson being home to the O’odham and the Yaqui. Committed to diversity and inclusion, the University strives to build sustainable relationships with sovereign Native Nations and Indigenous communities through education offerings, partnerships, and community service.

This work made use of the following software: \textsc{numpy} \citep{harris2020}; \textsc{matplotlib} \citep{hunter2007}; \textsc{scipy} \citep{virtanen2020}; \textsc{astropy}\footnote{\url{https://www.astropy.org/}}, a community-developed core Python package for Astronomy \citep{astropy2013, astropy2018}; \textsc{Source Extractor} \citep{bertin1996} via \textsc{sep} \citep{barbary2016}; \textsc{photutils} \citep{bradley2020}; \textsc{beagle} \citep{chevallard2016}; \textsc{multinest} \citep{feroz2008, feroz2009, feroz2019};  \textsc{prospector} \cite{johnson2021}; \textsc{dynesty} \citep{speagle2020}; \textsc{sedpy} \citep{johnson_sedpy}; and \textsc{fsps} \citep{conroy2009, conroy2010} via \textsc{python-fsps} \citep{johnson_python_fsps}


\section*{Data Availability}

The \textit{HST}/ACS and \textit{JWST}/NIRCam images used in this work are available through the Mikulski Archive for Space Telescopes (https://mast.stsci.edu/) under Proposal ID 1345. Additional data products and analysis code will be made available upon reasonable request to the corresponding author.



\bibliographystyle{mnras}
\bibliography{refs}



\appendix

\section{LAE-targeted selection} \label{appendix:lae_selection}

In Section\ \ref{subsec:selection}, we describe a Lyman break selection designed to photometrically identify galaxies at $z \sim 8.3 - 9.2$ as partial F115W dropouts. However, the presence of strong \Lya\ in a galaxy's spectrum at $z \sim 8.7$ will boost the F115W flux, making the $\text{F115W} - \text{F150W}$ colour bluer at a given redshift. This may have the effect of removing objects with the strongest \Lya\ emission ($\text{EW}_\Lya \gtrsim 75$\,\AA) from our selection at our targeted redshifts. Furthermore, we may expect galaxies in overdensities powering ionized bubbles to emit strong \Lya. Thus, to ensure our conclusions do not rely strongly on this potential bias against the strongest LAEs at $z \sim 8.7$, we supplement our primary selection with one designed to identify strong Ly$\alpha$ emitters ($\text{EW}_\Lya \sim 20 - 50$\,\AA) at $z \sim 8.1 - 9.5$:
\begin{enumerate}
    \item $\text{S/N} < 2$ in F435W, F606W, and F814W
    \item $\chi^2_\text{opt} < 5$
    \item $\text{F606W} - \text{F115W} > 1.2$ and $\text{F814W} - \text{F115W} > 1.2$
    \item $\text{F606W} - \text{F150W} > 1.7$ and $\text{F814W} - \text{F150W} > 1.7$
    \item $\text{F115W} - \text{F150W} > 0.3$ and $\text{F115W} - \text{F150W} < 1.4$
    \item $\text{F150W} - \text{F277W} < 0.6$
    \item $\text{F356W} - \text{F444W} > 0$
    \item $\text{F410M} - \text{F444W} > 0.1$
\end{enumerate}
We note that we add the additional requirement of a flux excess in F444W, as we expect strong Ly$\alpha$ emission to correlate with strong \ion{[O}{iii]}+\ion{H}{$\beta$} emission, which transmits through F444W at the redshifts of interest.

Applying this selection to our catalog and conducting the same visual inspection as described in Section\ \ref{subsec:selection}, we identify 13 additional candidates, summarized in Table\ \ref{tab:lae_selection}. We also calculate the increase in the surface density over the field, assuming all of these objects are LAEs at redshifts around $z \sim 8.7$. We find that the addition of these objects to our fiducial sample increases the surface density in all distance bins by $0.1 - 0.2$\,arcmin$^{-2}$, resulting in a surface density of $\sim (0.55 \pm 0.14)$\,arcmin$^{-2}$ within $\sim 5$\,arcmin of EGSY8p7. However, we caution that a detailed interpretation of the implied overdensity would require characterization of the completeness of our LAE-targeted selection. Furthermore, if these objects are not LAEs, the very weak $\text{F115W} - \text{F150W}$ break that we require would imply they lie at lower redshifts than the two known bright LAEs.

\begin{table*}
    \renewcommand{\arraystretch}{1.5}
    \centering
    \caption{The observed properties of the 13 new objects identified by our LAE-targeted selection sorted by increasing angular distance from EGSY8p7. We note that a very red $\text{F410M} - \text{F444W}$ colour empirically suggests strong \ion{[O}{iii]} and/or \ion{H}{$\beta$} emission transmitting through the F444W filter (a colour of $\text{F410M} - \text{F444W} = 0.5$ approximately corresponds to an equivalent width of $\text{EW} \approx 650$\,\AA\ at $z = 8.7$), which we may expect to correlate with strong \Lya\ emission.}
    \label{tab:lae_selection}
    \begin{tabular}{c|c|c|c|c|c}
    \hline
    \multirow{2}{*}{ID} & \multirow{2}{*}{RA} & \multirow{2}{*}{Dec} & Distance from & \multirow{2}{*}{F200W} & \multirow{2}{*}{$\text{F410M} - \text{F444W}$} \\
    & & & EGSY8p7 [\arcmin] & & \\ \hline\hline
    74335 & 14:20:11.21 & +52:53:59.94 & 0.7 & $26.7_{-0.1}^{+0.1}$ & $0.3_{+0.2}^{+0.2}$ \\
    49184 & 14:19:55.00 & +52:51:15.80 & 3.0 & $27.2_{-0.1}^{+0.1}$ & $0.1_{+0.2}^{+0.2}$ \\
    95370 & 14:19:41.58 & +52:55:09.54 & 4.4 & $28.0_{-0.1}^{+0.1}$ & $0.5_{+0.2}^{+0.2}$ \\
    52068 & 14:19:36.12 & +52:51:32.22 & 5.2 & $27.3_{-0.1}^{+0.1}$ & $0.6_{+0.3}^{+0.4}$ \\
    129391 & 14:20:33.81 & +52:57:04.72 & 5.3 & $26.9_{-0.1}^{+0.1}$ & $1.1_{+0.2}^{+0.2}$ \\
    109640 & 14:20:07.47 & +52:59:22.87 & 5.9 & $27.7_{-0.1}^{+0.1}$ & $0.5_{+0.2}^{+0.2}$ \\
    45859 & 14:19:22.75 & +52:50:51.27 & 7.4 & $27.5_{-0.1}^{+0.1}$ & $0.6_{+0.2}^{+0.2}$ \\
    48090 & 14:19:20.63 & +52:51:08.04 & 7.6 & $27.2_{-0.1}^{+0.1}$ & $0.6_{+0.3}^{+0.3}$ \\
    18931 & 14:19:34.53 & +52:47:26.89 & 7.9 & $27.5_{-0.1}^{+0.1}$ & $0.3_{+0.1}^{+0.2}$ \\
    37216 & 14:19:08.61 & +52:49:53.02 & 9.7 & $26.9_{-0.1}^{+0.1}$ & $0.2_{+0.2}^{+0.2}$ \\
    47548 & 14:19:05.24 & +52:51:03.45 & 9.8 & $27.7_{-0.1}^{+0.1}$ & $0.4_{+0.1}^{+0.1}$ \\
    25263 & 14:19:08.68 & +52:48:19.34 & 10.4 & $27.0_{-0.1}^{+0.1}$ & $0.3_{+0.2}^{+0.2}$ \\
    6586 & 14:18:47.59 & +52:45:05.55 & 14.8 & $26.3_{-0.0}^{+0.0}$ & $0.7_{+0.1}^{+0.1}$ \\ \hline
    \end{tabular}
\end{table*}

\section{Objects removed during visual inspection} \label{appendix:visual_inspection}

As described in Section\ \ref{subsec:selection}, we have visually inspected all colour-selected candidates and removed objects that are significantly impacted by an artifact (e.g. diffraction spikes and `snowballs' caused by significant cosmic ray events), appear to have unreliable background estimates, or appear to have visible flux in any of the three ACS optical filters. We have removed 20 objects from our fiducial sample, summarized in Table\ \ref{tab:removed_objects}.

\begin{table*}
    \renewcommand{\arraystretch}{1.5}
    \centering
    \caption{The objects that we removed during visual inspection. We report their coordinates and the reason for removal from our fiducial sample.}
    \label{tab:removed_objects}
    \begin{tabular}{c|c|c|c}
        \hline
        ID & RA & Dec & Reason for removal \\ \hline\hline
        9727 & 14:18:56.99 & +52:45:43.07 & Tentative ACS flux \\
        15604 & 14:19:32.35 & +52:46:56.68 & Unreliable background (F115W, F150W), snowball (F115W) \\
        18327 & 14:19:32.25 & +52:47:22.15 & Diffraction spike \\
        18350 & 14:19:32.23 & +52:47:22.27 & Diffraction spike \\
        21057 & 14:19:04.63 & +52:47:44.65 & Tentative ACS flux \\
        25508 & 14:19:29.32 & +52:48:21.17 & Photometric excesses implying $z \sim 2$ extreme emission lines, no acceptable $z \geq 6$ solutions from SED models \\
        35357 & 14:19:05.88 & +52:49:39.35 & Tentative ACS flux, no compact emission in F410M and F444W  \\
        41322 & 14:19:39.97 & +52:50:20.94 & Diffraction spike \\
        41349 & 14:19:40.00 & +52:50:21.25 & Diffraction spike \\
        46669 & 14:19:55.97 & +52:50:57.03 & F150W contaminated by diffraction spike \\
        48106 & 14:19:56.16 & +52:51:08.36 & Snowball (F150W) \\
        48545 & 14:19:16.78 & +52:51:11.34 & Tentative ACS flux (possibly from close neighbour) \\
        52131 & 14:19:20.15 & +52:51:32.20 & Tentative ACS flux \\
        62173 & 14:19:58.56 & +52:52.34.60 & Tentative ACS flux \\
        67680 & 14:19:37.73 & +52:53:11.16 & No clear compact emission \\
        70814 & 14:20:12.95 & +52:53:34.97 & Spurious detection \\
        87768 & 14:19:38.77 & +52:56:01.08 & Unreliable background (F115W, F150W) \\
        110426 & 14:20:06.50 & +52:59:17.41 & Unreliable background (F115W) \\
        118247 & 14:19:44.72 & +52:58:20.54 & Unreliable background (F115W, F150W, F200W) and near detector edge (F277W, F356W) \\
        124043 & 14:19:47.43 & +52:57:43.26 & Diffraction spike \\ \hline
    \end{tabular}
\end{table*}

\section{The impact of clumping factor on ionized bubble sizes} \label{appendix:clumping_factor}

\begin{figure*}
    \centering
    \includegraphics[width=\textwidth]{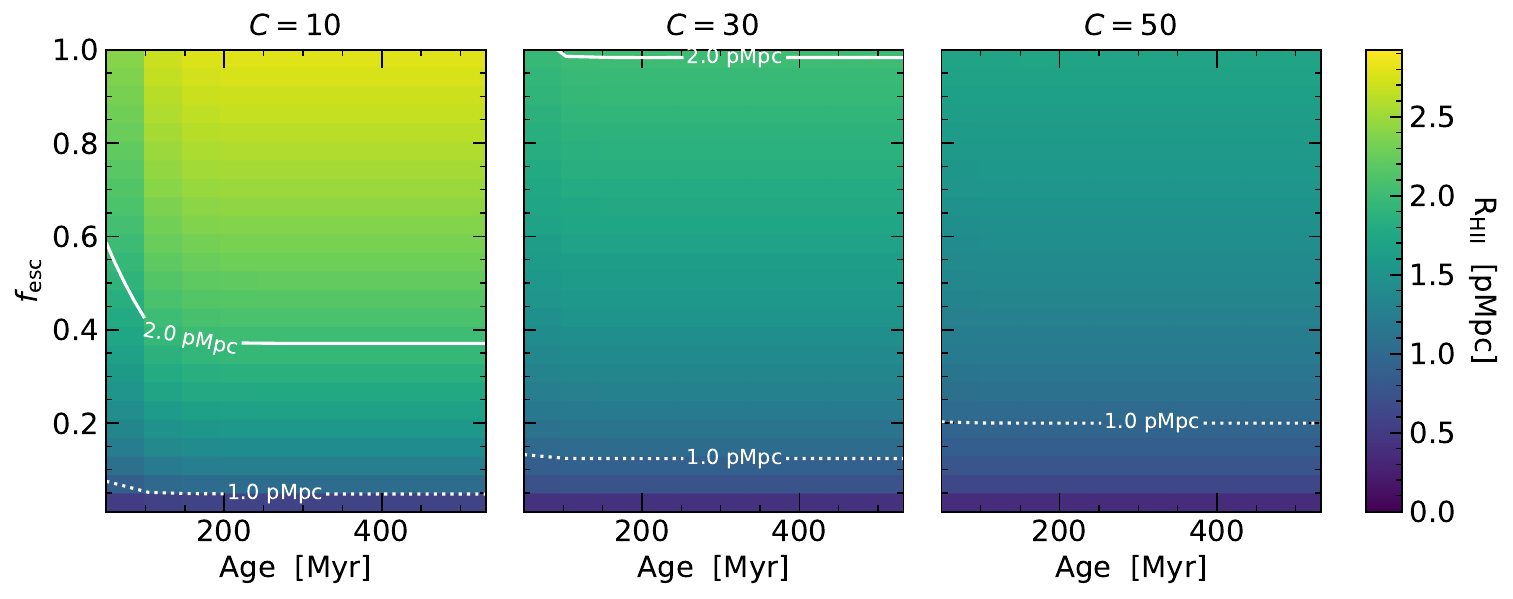}
    \caption{Predicted sizes of the ionized bubble that could be created by a four times overdense population of $M_\textsc{uv} \leq -15$ galaxies within a sphere of radius $R = 2$\,pMpc. We show three different IGM clumping factors (left: $C = 10$, center: $C = 30$, right: $C = 50$). To ionize an $R \geq 2$\,pMpc spherical volume with larger clumping factors, galaxies require increasingly large Lyman continuum escape fractions over long times due to increased recombination rates.}
    \label{fig:R_HII_clumping}
\end{figure*}

In Section\ \ref{sec:overdensity}, we estimate the size of the ionized bubble that could be created by a population of $M_\textsc{uv} \leq -15$ galaxies as a function of escape fraction, $f_\text{esc}$ and ionizing photon production emissivity, $\xi_\text{ion}$. We balance the rate of growth due to photoionizations and the Hubble flow and the recombination rate in the IGM (Equation\ \eqref{eqn:dR_dt}), assuming an average IGM `clumping factor' ($C = \langle n_{\ion{H}{i}}^2 \rangle / \langle n_{\ion{H}{i}} \rangle^2$) of $C = 3$, consistent with expectations from simulations for the average IGM \citep[e.g.][]{shull2012, finlator2012, kaurov2015, gorce2018}. However, if the region under consideration in this work is overdense, the clumping factor may be larger. We test the impact of adopting larger clumping factors \citep[$C = 10, 30, 50$;][]{mao2020, bianco2021} on the final ionized bubble sizes; see Figure\ \ref{fig:R_HII_clumping}. Due to the increased recombination rate in the IGM, larger clumping factors lead to smaller bubble sizes at fixed escape fraction, though the population of galaxies under consideration can still power an $R \geq 2$\,pMpc ionized bubble with $f_\text{esc} \sim 0.4$ \citep[consistent with some individual observations of Lyman continuum leakers at lower redshift,][]{izotov2018, flury2022}, over hundreds of Myr.

\bsp	
\label{lastpage}
\end{document}